\documentclass[sigconf,nonacm]{acmart}
\AtBeginDocument{%
  }

\newcommand{\point}[1]{%
  \vspace{0.5em} 
  \noindent 
  \textbf{$\diamond$ \textit{#1}}
}

\usepackage{xspace}
\newcommand{\tool}{\textit{Inkitiate}\xspace}

\newcommand{\workicon}[1]{%
  \raisebox{-0.20\height}{%
    \includegraphics[width=0.34cm]{figures/icons/#1.pdf}%
  }\nobreak\hspace{0.15em}%
}

\newcommand{\bringmaterials}{\workicon{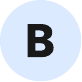}}
\newcommand{\thinkwithmaterials}{\workicon{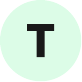}}
\newcommand{\shapethespace}{\workicon{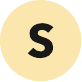}}

\usepackage{xcolor}

\usepackage{tikz}

\definecolor{capabilitySigil}{HTML}{6956C7}
\definecolor{capabilitySigilBG}{HTML}{EFEDFF}
\definecolor{resourceSigil}{HTML}{2E8B63}
\definecolor{resourceSigilBG}{HTML}{E8F5ED}

\newcommand{\sigilmark}[3]{%
  \tikz[baseline=(sigil.base)]{%
    \node[
      fill=#1,
      rounded corners=1.2pt,
      inner xsep=2pt,
      inner ysep=0pt,
      minimum height=1.35em,
      text height=1.05ex,
      text depth=0.20ex
    ] (sigil) {%
      \textcolor{#2}{%
        {#3}%
      }%
    };
  }%
}

\newcommand{\resourcesigil}{%
  \sigilmark{resourceSigilBG}{resourceSigil}{\#}%
}

\newcommand{\capabilitysigil}{%
  \sigilmark{capabilitySigilBG}{capabilitySigil}{//}%
}

\begin{document}

\title[Inkitiate]{Inkitiate: Initiating and Steering Agent Assistance through Ink in a Spatial Workspace}

\author{Liangwei Wang}
\orcid{0000-0003-3481-3993}
\affiliation{
  \institution{HKUST(GZ)}
  \country{Guangzhou, China}
}
\email{lwang344@connect.hkust-gz.edu.cn}

\author{Nan Chen}
\orcid{}
\affiliation{
  \institution{Microsoft Research Asia}
  \country{Shanghai, China}
}
\email{nanchen@microsoft.com}

\author{Luna K. Qiu}
\affiliation{%
  \institution{Microsoft Research Asia}
  \city{Shanghai}
  \country{China}}
\email{lunaqiu@microsoft.com}

\author{Zilong Wang}
\orcid{}
\affiliation{
  \institution{CataX AI}
  \country{Shanghai, China}
}
\email{wangzilong@cata-x.ai}

\author{Fugee Tsung}
\orcid{0000-0002-0575-8254}
\affiliation{
  \institution{HKUST}
  \country{Hong Kong SAR, China}
}
\email{season@ust.hk}

\author{Yuyu Luo}
\orcid{0000-0001-9530-3327}
\affiliation{
  \institution{HKUST(GZ)}
  \country{Guangzhou, China}
}
\email{yuyuluo@hkust-gz.edu.cn}

\author{Yuqing Yang}
\orcid{}
\affiliation{
  \institution{Microsoft Research Asia}
  \country{Shanghai, China}
}
\email{yuqyang@microsoft.com}

\renewcommand{\shortauthors}{Wang et al.}

\begin{abstract}
Handwriting and sketching help people develop ideas around information.
As this work unfolds, users may seek assistance with finding, understanding, and organizing materials.
We investigate how ink can initiate and steer agent assistance in this process.
An eight-participant formative study explored desired assistance and interaction expectations, informing three connected activities: bringing materials in, thinking with materials, and shaping the space.
We present \tool, a spatial workspace for ink-based agent assistance in knowledge work.
Through ink, users retrieve external materials, build on accumulated content and results, and reorganize the canvas, while inspecting and guiding agent actions.
In a study with eighteen participants, users reported favorable usability and low perceived workload, and used retained ink and materials to develop further ink queries.
An ink-query vocabulary and cases reveal how users combine linguistic and graphical expressions to direct agents.
These findings inform ink-based agent interaction design for expressive requests and continued work with materials.
\end{abstract}




\begin{teaserfigure}
  \centering
  \includegraphics[width=\linewidth]{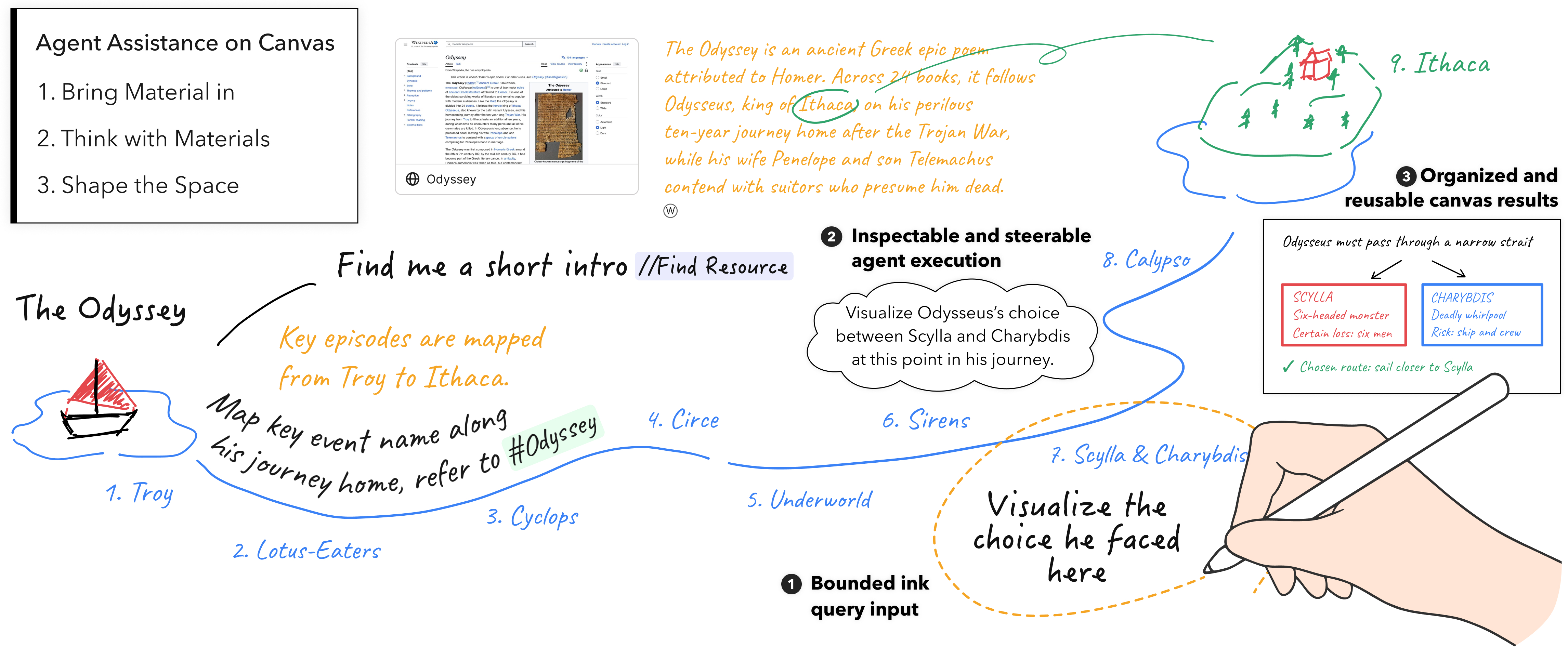}
  \caption{
  Inkitiate is a spatial knowledge workspace where users interact with diverse canvas objects and use ink expressions to initiate and steer agent assistance directly on the canvas. Here, Amy explores Odysseus’s journey from Troy to Ithaca by working with web resources, text, drawings, and spatial paths. She (1) creates a bounded query from a selected ink expression, (2) inspects and steers the agent’s execution, and (3) incorporates its organized, reusable results into her evolving canvas.}
  \vspace{1.2em}
  \Description{A hand-drawn spatial canvas shows Amy exploring the Odyssey with an agent. On the left, a boat marks Troy beside the handwritten request “Find me a short intro,” a web resource card for the Odyssey, and a note asking to map key events along Odysseus’s journey home. A blue path labels episodes from Troy through the Lotus-Eaters, Cyclops, Circe, Underworld, Sirens, Scylla and Charybdis, and Calypso to Ithaca. Near the center, Amy circles the request “Visualize the choice he faced here,” forming a bounded ink query. An agent card presents the interpreted task and lets her inspect and steer execution. On the right, the canvas contains a map of Ithaca and an organized comparison of Scylla and Charybdis as reusable results. A panel identifies three forms of assistance: bringing materials in, thinking with materials, and shaping the space.}
  \label{fig:teaser}
\end{teaserfigure}

\maketitle

\section{Introduction}

Handwriting and sketching help people externalize and develop ideas that are still taking shape~\cite{riche2017aswemayink}. In digital interfaces, these handwritten and sketched marks---digital ink---can serve not only as notes or illustrations but also as input for computational tasks. Generative AI systems use sketches to guide the creation of visual content, including images and animations~\cite{marquardt2025imaginationvellum,li2026sketchdynamics}. Ink can also be interpreted in context to support code editing, document feedback, and tasks across applications~\cite{yen2025codeshaping,leung2026annotategpt,huang2025sketchgpt}. These examples illustrate how ink can express both what users want to create and what they want a system to do with existing content.

Knowledge work often involves sensemaking: an iterative process of gathering information and developing and revising interpretations~\cite{russell1993cost,pirolli2005sensemaking}. Digital spatial workspaces support sensemaking by enabling users to retain, group, and rearrange documents and notes, keeping relevant information available and expressing relationships as their understanding develops~\cite{kidd1994marks,andrews2010spacetothink,andrews2012analystworkspace}. With large language models, these workspaces also support AI-assisted exploration, synthesis, and organization, whose results can become materials for subsequent tasks~\cite{suh2023sensecape,kang2023synergi,choe2026crosslit}.

Within these workspaces, users seek agent assistance with questions and ideas expressed in their notes, sketches, and annotations, using relevant materials as context. 
The results become part of the workspace for users to examine, reorganize, and build upon as their understanding develops. 
However, ink rarely specifies a complete task on its own: its linguistic, visual, spatial, and relational information must be interpreted together with selected materials and prior results to determine the task's referents, scope, and intended action.
As an agent interprets and executes such a situated request, users also need to understand which materials and interpretations are shaping the task and be able to clarify, correct, or stop it when execution diverges from their intent.
We therefore ask: \emph{How can ink interaction enable users to seek and steer agent assistance within a spatial workspace to advance ongoing knowledge work?}

We began with a formative study involving 8 participants and an initial pen-first workspace prototype. The prototype served as a design probe for exploring where participants wanted agent assistance and how they expected to interact with it. The study highlighted a need for cumulative agent assistance, in which users revisit and reorganize existing materials and use prior results to develop subsequent inquiries, rather than treating each ink request as an isolated question--answer exchange. 
Participants expected their selections and spatial marks to delimit the content the agent should use and indicate where its results should appear. When the prototype failed to meet these expectations, participants wanted ways to inspect and correct its interpretation and stop unwanted actions. 
Drawing on these insights, we derived key design implications that guided our system development.

We present \tool\footnote{Inkitiate is implemented as a feature within Huabu, a broader open-source spatial workspace. The interaction design and features described in this paper are available in the Huabu repository: \url{https://github.com/microsoft/Huabu}.}, a spatial workspace in which users initiate and guide agent assistance through ink to bring materials in, think with them, and shape the space. Each task begins with an explicit submission of ink, optionally with selected canvas objects, which the system interprets using bounded spatial context and relevant history from prior ink queries. During execution, the interface exposes the agent's understanding, included materials, and progress. Users can respond to clarification questions through suggested options or additional ink, and abort execution when needed. The system writes results back to the canvas by creating or modifying persistent, editable objects. These objects remain available for reorganization and reuse in subsequent ink queries, allowing agent-assisted work to build on what is already in the workspace.

We examined \tool with 18 participants through prescribed knowledge-work activities (\emph{Reproduction}) and self-directed exploration (\emph{Freeform Creation}). Participants reported favorable usability and low perceived workload. Their experiences highlighted the reuse of retained ink and materials, difficulties addressing local targets, and the importance of output form for further work. An ink query vocabulary characterizes how participants combined linguistic, referential, and spatial marks to express their intentions, while qualitative cases illustrate ink's potential for guiding iterative image creation and precise revision, as well as the organization of materials into meaningful visual structures. Together, these observations inform our discussion of ink's value for directing agents and of designs that support continued work with materials while preserving user direction. Our contributions are:
\begin{itemize}
	\item A formative account of three connected activities for ink-initiated agent assistance--bringing materials in, thinking with materials, and shaping the space--along with expecting interaction and the resulting design goals.
	\item \tool, a spatial workspace that connects explicit ink submission with contextual task interpretation, user intervention during execution, and persistent, reusable results.
	\item An empirical account of how people use ink to direct agents, characterizing an ink-query vocabulary and illustrating interaction possibilities through cases.
\end{itemize}

\section{Related Work}

To contextualize \tool's design and the challenges of ink-initiated agent assistance, we review related work across three areas.

\subsection{Ink as Content and Instruction}

Handwriting and sketching provide a flexible medium for externalizing ideas and annotating existing materials~\cite{riche2017aswemayink,kim2019inking}.
In digital systems, ink can serve both as content to interpret and augment and as a means of indicating what users want the system to do, with its meaning shaped by its form, location, and relationship to surrounding materials~\cite{shi2024inkeraction,yen2025codeshaping}.

Ink externalizes users' ideas, making them available for later review and computational interpretation.
Research on data analysis found that free-form inking supported graphical externalizations and that explicit links between ink marks and data helped users revisit their work~\cite{kim2019inking}.
AmIWrite~\cite{liu2026amiwrite} interprets handwritten mathematical work to provide feedback.
Inkeraction~\cite{shi2024inkeraction} recognizes how ink is grouped and connected, adjusting layout and preserving connections as users rearrange content.
Sketches also provide visual content for generation: ImaginationVellum~\cite{marquardt2025imaginationvellum} uses sketches as spatial prompting for image generation, Inkspire~\cite{lin2025inkspire} supports iterative product design, and SketchDynamics~\cite{li2026sketchdynamics} interprets sketches as dynamic behavior for animation.

Ink can also express instructions, with its content and context helping systems determine the intended actions and their targets.
texSketch~\cite{subramonyam2020texsketch} uses ink annotations made while reading to guide causal diagram construction, and ActiveInk~\cite{romat2019activeink} lets users activate existing ink annotations as predefined analytic operations.
InkSight~\cite{lin2024inksight} uses sketch-based selections to identify chart data for generating findings and documentation.
SketchingRelatedWork~\cite{ma2023sketchingrelatedwork} maps hand-drawn boxes, handwriting, and node-link structures onto paper and citation queries.
For free-form requests, Code Shaping~\cite{yen2025codeshaping} and Notational Animating~\cite{shi2026notational} interpret contextual sketches to guide code edits and animation keyframes, respectively;
SketchGPT~\cite{huang2025sketchgpt} combines sketch and speech to express task intentions across existing application interfaces.
Thinkink~\cite{payandeh2026thinkink} supports iterative ideation around handwritten notes, sketches, and model responses, allowing users to formulate requests using canvas content and revise generated content through ink annotations.
These studies demonstrate how ink can both externalize ideas and communicate intended actions.
They motivate us to explore how users can draw on these roles to initiate and guide agent assistance in knowledge work.

\subsection{Evolving Workspaces for Knowledge Work}

Sensemaking in knowledge work involves repeatedly gathering information, developing interpretations, and revising understanding as new materials emerge.
Foundational work frame sensemaking as constructing and reusing external representations: \citet{russell1993cost} analyze its cost structure, and \citet{pirolli2005sensemaking} distinguish an information-foraging loop from a sensemaking loop with evidence, schema, and hypothesis stages.
Analytic workspaces support this process by keeping documents and evidence available for inspection, organization, and manipulation~\cite{wright2006sandbox,stasko2008jigsaw,bier2008entity}.
In such workspaces, large displays can further support working memory and externalize evolving analytic structures through spatial layout~\cite{andrews2010spacetothink,andrews2012analystworkspace}.

Recent systems integrate LLM assistance into workspaces where users can revisit and build on earlier materials and results.
Graphologue~\cite{jiang2023graphologue} turns model responses into interactive node-link diagrams, while Sensecape~\cite{suh2023sensecape} supports exploration across levels of abstraction.
Memolet~\cite{yen2024memolet} addresses context reuse by turning conversational memories into manipulable objects that users curate, reference, and adjust to guide subsequent generations.
IntentTagger~\cite{intenttagging2025gmeiner} externalizes generation intent as small, composable tags, including references to existing content, which users can refine to steer deck-level generation and subsequent slide-level edits.
In scholarly and qualitative knowledge work, Garden of Papers~\cite{ma2025gardenofpapers} connects paper finding, reading, and organization, while ScholarMate~\cite{ye2025scholarmate} explores mixed-initiative sensemaking.
CrossLit~\cite{choe2026crosslit} further links the spatial organization of papers and notes with editable textual structures, allowing users to revise an emerging review across visual and textual representations.
Contemporary products such as Heptabase~\cite{heptabase2026}, Miro~\cite{miro2026canvascontent}, tldraw~\cite{tldraw2026}, and Microsoft's Huabu~\cite{microsoft2026huabu} also provide persistent spatial workspaces for organizing, connecting, and revisiting heterogeneous materials.

Prior work therefore shows how spatial workspaces can preserve and reorganize materials and intermediate results as users’ understanding develops. 
Complementing these systems, Thinkink~\cite{payandeh2026thinkink} supports iterative ideation through user ink and LLM-generated responses on a shared canvas. 
However, it remains underexplored how situated ink can initiate and steer agents across a broader knowledge-work process: acquiring new information, reasoning over selected materials, and creating or reorganizing workspace content. 
As these actions operate over heterogeneous, source-bearing materials and prior results, users must communicate not only what they want but also which objects the agent should act on and how. 
We investigate how situated ink can support this form of agent assistance in an evolving workspace.

\subsection{Interaction-Augmented Instructions and User Control}

Users can communicate their intentions to AI systems through interactions with visible materials as well as language. 
Earlier work on semantic interaction explored how users' manipulation of analytic representations can reveal their intent and guide updates to underlying computational models~\cite{endert2012semantic}. 
For generative AI, interaction-augmented instructions combine GUI actions, language, and artifact context to form the instruction received by the model~\cite{shen2026iai}.

When systems infer part of a request from these interactions, users need ways to inspect and correct the resulting interpretation and its effects.
Guidelines for human--AI interaction similarly emphasize helping users understand system behavior and supporting efficient correction~\cite{amershi2019guidelines}.
S-PRISM~\cite{tang2026semanticprompting} interprets spatial interactions as report-refinement intentions, presents these interpretations through interface bubbles, and highlights changes to the report.
DirectGPT~\cite{masson2024directgpt}
lets users select visible objects for model operations and undo edits,
while VizTA~\cite{wang2025vizta} connects references to chart elements with conversational explanations.
AI Chains~\cite{wu2022aichains} exposes model calls and intermediate outputs as editable modules, allowing users to inspect and debug individual steps.
DynaVis~\cite{vaithilingam2024dynavis} synthesizes persistent visualization-editing controls from natural language requests, letting users adjust results directly rather than restate each change as a new request.

User involvement also needs to continue when execution encounters missing information or decisions that depend on user preferences~\cite{shen2026iai}.
Morae~\cite{peng2025morae}, designed for blind and low-vision users, pauses at decision points where preferences are unclear and elicits choices through generated interfaces.
CowPilot~\cite{huq2025cowpilot} supports user-initiated intervention: users can pause or override agent actions, act directly, and subsequently return control to the agent.
In research tasks, ResearStudio~\cite{yang2025researstudio} exposes a live plan and action history, allowing users to pause execution, edit the plan or code, and resume the task.
Recent research on computer-use-agent interfaces likewise highlights that effective user control depends on making agent behavior sufficiently understandable to support informed intervention during execution~\cite{cheng2026mapping}.

Together, this work shows how interaction can help form AI instructions and support inspection and intervention throughout execution.
Yet situated ink requests combine free-form marks, spatial relations, selected materials, and evolving workspace context, while agents may take multi-step actions that alter the workspace.
How to keep such tasks bounded, inspectable, and steerable remains underexplored; we examine how users can submit ink with relevant materials, inspect how agents interpret and execute these requests, and intervene as the task unfolds.

\section{Design Foundations}
\label{sec:design-foundations}

We conducted a formative study to understand how users would use ink to seek and guide agent assistance in ongoing knowledge work, and what interaction support they would need.

\subsection{Formative Study}

\subsubsection{Initial Prototype}

\begin{figure*}[t]
  \centering
  \includegraphics[width=\linewidth]{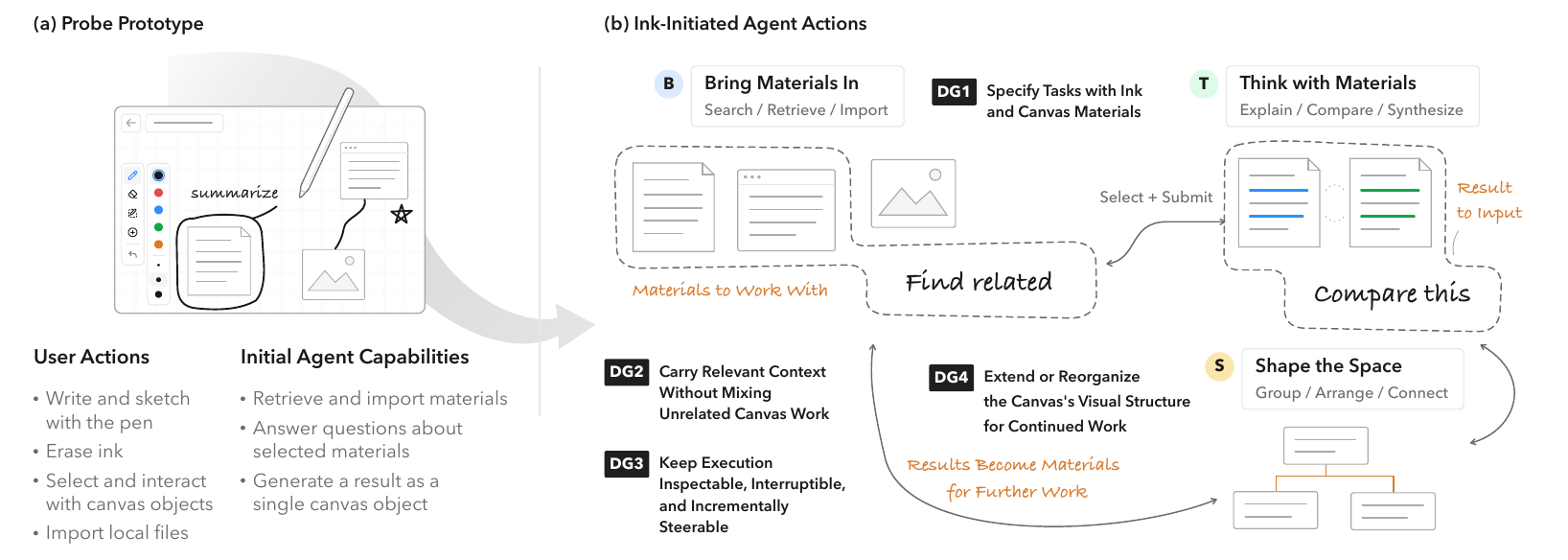}
  \caption{From the initial probe to ink-initiated agent interaction. (a) The probe enabled users to use ink and canvas objects to seek and guide agent assistance. (b) Prototype use and participant feedback revealed three recurring activities—bringing materials in, thinking with materials, and shaping the space—and four interaction expectations that informed DG1–DG4.}
  \Description{A two-panel diagram showing how the initial probe informed the design of ink-initiated agent actions. Panel (a) depicts the probe's spatial canvas interface and summarizes its capabilities. Users could write and sketch with ink, erase ink, select and interact with canvas objects, and import local files. The agent could retrieve or import materials, answer questions about selected materials, and return a generated result as a canvas object. Panel (b) organizes the resulting interaction design around three connected activities. Bringing Materials In uses ink and selected canvas materials to retrieve or generate resources. Thinking with Materials supports explanation, comparison, and synthesis using existing materials. Shaping the Space uses ink to group, arrange, and connect canvas objects. Arrows indicate that results return to the canvas as materials for continued work. Four design goals address specifying tasks with ink and canvas materials, carrying forward relevant context, keeping agent execution inspectable and steerable, and extending or reorganizing the canvas's visual structure.}
  \label{fig:scholarly-sensemaking}
\end{figure*}

We based the initial probe on the sensemaking model summarized in Figure~\ref{fig:scholarly-sensemaking}(a), focusing on the interplay between information foraging and the development of representations~\cite{pirolli2005sensemaking}.
Building on the information-foraging loop, the prototype provided a spatial canvas where users could use ink to steer the agent to retrieve external resources or generate content, bringing these materials into the canvas as text, images, webpages, or PDF objects.
To support sensemaking with these materials, users could then use ink to ask questions about existing objects or indicate intended changes.
After users submitted an ink query, the agent returned results as new canvas objects or updates to existing content once the run finished, making them available for further exploration and organization.

\subsubsection{Method}

We recruited 8 participants (P1--P8; 3 male, 5 female) with prior experience and an interest in digital handwriting.
Participants were recruited through convenience sampling at our research institute.
Formative sessions lasted about one hour, and participants each received compensation equivalent to 9 USD.

At the beginning of each session, we introduced the project, explained how to use the prototype, and invited participants to explore open-ended tasks based on their own interests.
Participants followed a think-aloud protocol during exploration, while we observed their interaction strategies, breakdowns, and unexpected system behaviors.
After the exploration, we interviewed participants about their experiences and expectations for ink-initiated agent assistance.
With participants' consent, we recorded screen activity and audio in all sessions and automatically transcribed the audio for analysis.
The first author performed open coding of the recordings and transcripts, iteratively refining the codes through discussions with a second author to derive the formative findings.

\subsubsection{Findings: Activities and Interaction Expectations}

Drawing on participants' use and feedback together with the sensemaking model, we identified three activities in which participants wanted ink-initiated agent assistance (Figure~\ref{fig:scholarly-sensemaking}(b)).
\emph{Bringing materials in} uses ink-initiated requests to retrieve or generate resources.
\emph{Thinking with materials} combines ink and materials on the canvas to seek explanations, comparisons, and synthesis.
\emph{Shaping the space} uses ink to guide the organization of materials and their relationships into editable visual structures.
These activities connect information foraging with interpretation and representation development, including the construction and revision of schemas.
They can overlap and recur as results become materials for further ink query.
Across these activities, participants expressed four expectations for ink-initiated agent interaction.

\point{Finding One: Participants wanted to combine brief ink expressions with spatial references and explicit object selection to specify their requests.}
Participants did not expect ink alone to carry a complete request.
P3 noted that the effort of handwriting would make their expressions shorter and potentially less explicit than typed requests.
Participants instead wanted to use compact ink to express what they wanted to do while relying on existing canvas objects to indicate what the request concerned.
For example, P5 suggested circling a location in a travel guide and writing \textit{``what shops are there?''}: the enclosure identified the referent, while the text specified the desired information.
Relying only on local spatial relationships, however, had limitations.
P5 noted that earlier material in a chat could become visually distant from the current interaction and difficult to refer back to; on a larger canvas, relevant objects likewise might not be located near the ink.
P8 wanted to explicitly select which objects were included when submitting an ink query and envisioned combining a bounded region within an object with a short instruction.
Together, these responses suggest that participants wanted to combine brief ink expressions with spatial references and explicit object selection to specify both the requested action and the materials it concerned.

\point{Finding Two: Participants expected context to follow the current line of work rather than either an individual query or the entire canvas.}
Participants wanted the agent to use the canvas materials and prior history relevant to the current task, but did not expect all canvas content to enter a single shared context.
P5 valued that context on the canvas could remain spatially organized rather than being arranged only chronologically and gradually pushed out of view, as in a chat history.
Preserving context, however, did not mean expanding it without limit.
When P4 asked beside a paper how to reproduce its 3D MRI pipeline, the system returned a general learning path; P4 felt that the relationship between the request and the paper had been lost.
Conversely, after a travel request produced an unexpected response, P7 suspected that the system had used too much context and argued that the current task required only the relevant guide material.
These interpretations do not establish the technical causes of the failures, but reveal expectations at both sides of the context boundary: context needed to preserve the objects, spatial relationships, and history required by the current line of work while excluding canvas content unrelated to the task.
The expected unit of context was therefore closer to an evolving line of work than to either a single ink query or the entire canvas.

\point{Finding Three: Participants wanted visibility into the agent's interpretation and progress, with opportunities to intervene during execution.}
In the initial probe, results appeared on the canvas only after an agent run finished, leaving users unable to inspect or intervene during execution.
Participants identified three related needs.
First, they needed basic awareness of whether the agent was still active.
When a precise paper search ran long without canvas feedback, P4 wanted a status indicator to \textit{``let me know it is working.''}
Second, participants wanted to correct a misinterpreted or misdirected action without discarding the larger task.
After an image request returned the wrong cat breeds, P8 wanted to see how the agent had interpreted her intent while it was still executing, so that she could interrupt execution or revise the submitted ink query before completion.
Third, when the agent needed information from a limited set of possibilities, participants preferred a lightweight way to supply it.
When the agent asked about visa type during a French visa task, P3 suggested presenting selectable options rather than requiring further handwriting.
Together, these responses show that participants needed more than execution status; they wanted a continuing path for understanding, correcting, stopping, and clarifying the task without starting over.

\point{Finding Four: Participants expected write-back to extend or reshape the canvas's visual structure for continued work.}
In the initial probe, results remained on the canvas as persistent objects, but persistence alone did not make them part of continued work.
Participants used prior results as materials for further work.
After inspecting a generated itinerary, P7 first restated an omitted departure constraint and asked the agent to improve the route, then used the revised result to explore food along the way and images of a particular dish.
This process involved both repairing the current result and using it as material for further exploration.
However, continued work also required results to fit the canvas's visual structure through both their placement and their form.
When an answer appeared far from the relevant mind-map branch, P3 wanted the system to expand near the question mark and continue the current line of thought.
P5 similarly did not want results to accumulate downward like a chat history, suggesting that their arrangement should reflect the canvas's nonlinear relationships.
When existing materials became cluttered, participants wanted the agent to help reorganize the canvas rather than merely add more content.
P6 emphasized the layout of the \textit{``entire canvas''} and wanted to manually refine it after the agent had organized it.
Together, these responses suggest that write-back should do more than insert objects: it should extend meaningful local relationships or reorganize cluttered materials into a clear, editable structure under user control.

\subsection{Design Goals}

These findings informed four design goals for supporting ink-initiated agent interaction across the three activities.

\point{DG1: Enable users to combine ink with canvas materials to specify tasks.}
Users should be able to combine ink with relevant canvas materials, without having to restate those materials in the ink query.
Canvas materials, including prior results, should therefore remain persistent, selectable, manipulable, and source-bearing resources available to both the user and the agent.
Each task should remain pen-first: submitted ink initiates the task, while selected or explicitly referenced materials provide its targets and context.
This allows users to refer to nearby materials through spatial ink and to distant materials through explicit references without relying on proximity alone.

\point{DG2: Carry forward relevant context without mixing unrelated canvas work.}
A short ink query may depend on objects, spatial relationships, and results from earlier tasks.
Related requests should therefore be able to carry forward the materials and history needed for continued work.
At the same time, content should not enter a task merely because it exists elsewhere on the same canvas.
The system should maintain bounded continuity: users can continue an existing line of work without restating its context, while unrelated activities remain separate.

\point{DG3: Keep agent execution inspectable, interruptible, and incrementally steerable.}
Submitting an ink query should not turn the task into an opaque process that users can assess only after completion.
Users should be able to inspect how the agent interpreted the request, which materials it is using, and what activity is underway.
They should also be able to stop or redirect an action without discarding the larger task established on the canvas.
When missing information would affect the task's direction, the system should pause and request focused clarification through selectable options or a short custom response.

\point{DG4: Support continued work by extending or reorganizing the canvas's visual structure.}
Results should return as persistent and editable canvas objects rather than terminal responses.
Their placement and form should follow the ongoing work: results should remain near the originating ink and materials or follow spatial constraints expressed through ink.
When a result contains multiple elements or relationships, it should use a content-appropriate visual structure rather than default to a detached block of long-form text.
Users should also be able to use ink to steer the agent in reorganizing a bounded set of existing materials, including arranging them along drawn trajectories, and then manually refine the resulting arrangement.
Provenance should remain attached as results are moved, edited, organized, or reused in subsequent ink queries.

\section{\tool}
\label{sec:system}

\tool is a spatial workspace where users guide agent assistance through free-form ink to advance ongoing knowledge work. Because this work involves continuous interaction with heterogeneous materials, \tool is designed to support an evolving, iterative process. It makes the agent's task interpretation inspectable, allows users to intervene during execution, and returns editable, source-bearing results directly to the persistent canvas.

\begin{figure*}[t]
  \centering
  \includegraphics[width=\linewidth]{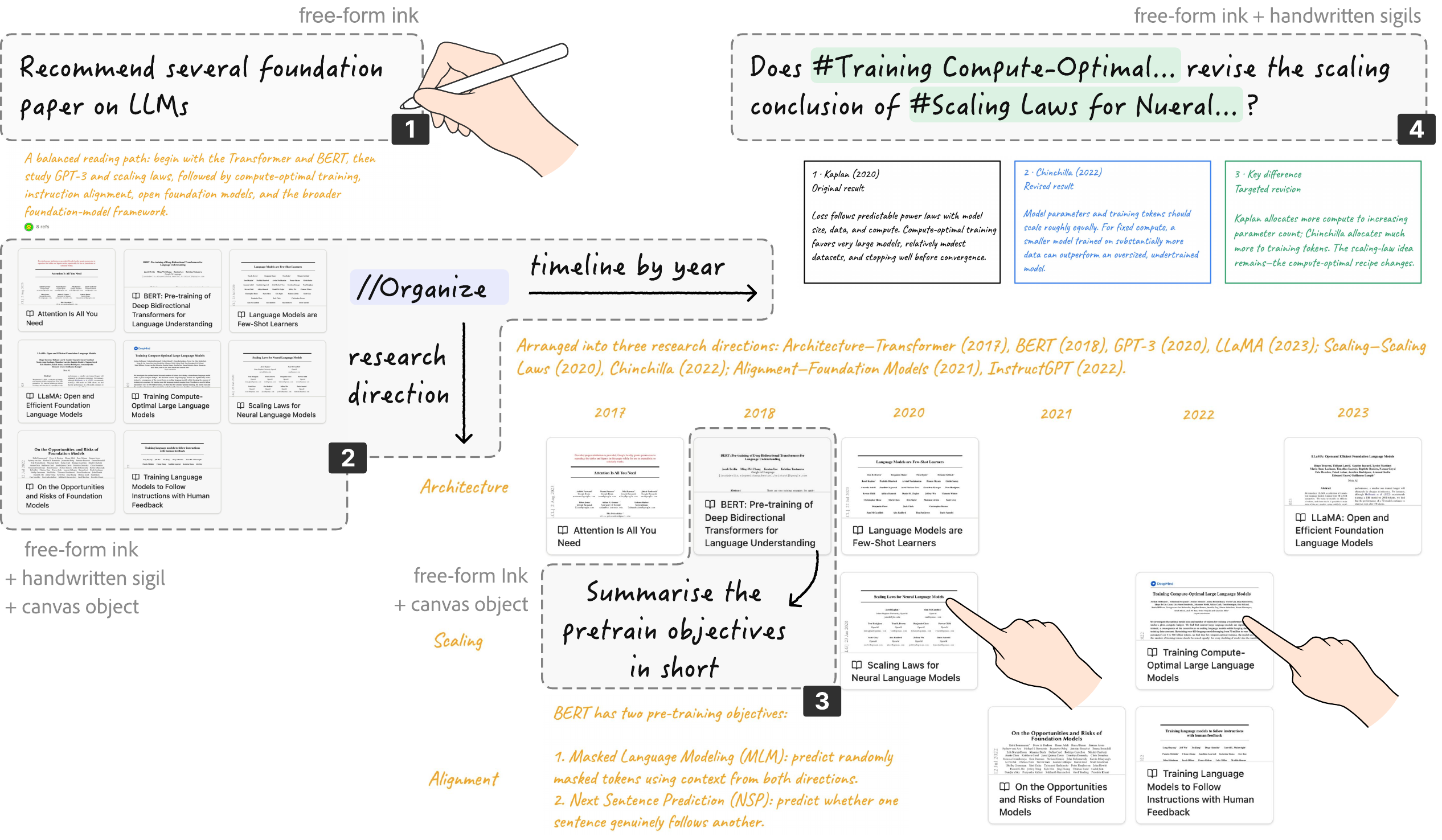}
  \caption{A bounded task input in \tool. In this example, the user lassos and explicitly submits free-form ink together with selected canvas objects and an optional sigil. Nearby materials outside the submission boundary remain on the persistent canvas but are not automatically included as task inputs.}
  \Description{Placeholder for a real \tool case showing a lasso boundary around submitted free-form ink, selected canvas objects, and an optional handwriting sigil. Other nearby canvas materials remain visibly outside the bounded submission.}
  \label{fig:bounded-task-input}
\end{figure*}

\subsection{System Walkthroughs}

We introduce \tool's interaction flow through two walkthroughs featuring a hypothetical user, Amy. 
Together, they demonstrate how sequences of ink queries help her advance ongoing knowledge work by \bringmaterials bringing materials in, \thinkwithmaterials thinking with them, and \shapethespace shaping their organization as her work develops.

\subsubsection{Exploring \textit{The Odyssey}.}
Amy first sketches a boat and writes \textit{``The Odyssey''} beside it, then draws an island and a route linking them to establish a spatial structure for the journey (Figure~\ref{fig:teaser}).
Beside the boat, she writes \textit{``Find me a short intro''} and adds a handwritten sigil \capabilitysigil{} to explicitly invoke resource retrieval.
\tool brings a source webpage and a concise, source-backed overview into the canvas.
Using the selection tool, Amy selects the curved route and its accompanying handwritten request, asking \tool to use the imported webpage \resourcesigil{} to map key episodes in Odysseus's journey from Troy to Ithaca along the route.
\tool places the episodes along the route, turning the initial sketch into a spatial account that remains available for further annotation and inquiry.
She then writes \textit{``Visualize the choice he faced here''} beside Scylla and Charybdis, lassos the ink expression, and submits it as a bounded query input (Fig.~\ref{fig:teaser}~1).
An agent activity card appears beside the query, exposing the agent's interpretation and current execution so that Amy can inspect and steer the task (Fig.~\ref{fig:teaser}~2).
As a result, \tool writes an organized comparison of Odysseus's alternatives back to the canvas as editable, reusable objects (Fig.~\ref{fig:teaser}~3).

\subsubsection{Reading foundational LLM papers.}
Amy begins a literature review in \tool by handwriting an ink query for several foundational papers on large language models (Fig.~\ref{fig:bounded-task-input}~1).
\tool brings the recommended papers into the canvas as PDF objects and provides a concise reading summary.
She then invokes the organize handwriting sigil \capabilitysigil{} and draws two labeled arrows asking the system to arrange the papers by publication year and research direction (Fig.~\ref{fig:bounded-task-input}~2).
\tool reorganizes the papers into a tabular literature overview that exposes both temporal and thematic relationships.
Amy next selects the BERT paper and submits a short ink query asking for a summary of its pretraining objectives (Fig.~\ref{fig:bounded-task-input}~3).
Finally, she writes the resource sigil \resourcesigil{} and selects two papers on scaling directly on the canvas to bind them to her request (Fig.~\ref{fig:bounded-task-input}~4).
She then asks how the later work revises the earlier conclusion, and \tool organizes the original conclusion, its revision, and the key difference between them into a structured result.
Across these steps, retrieved materials and agent-generated results remain available for subsequent ink queries and reorganization.

\subsection{Interaction Support for Inspecting and Steering Ink-Initiated Agent Tasks}

This section focuses on the lifecycle of a single agent task, detailing how \tool keeps it bounded, inspectable, and steerable from ink query submission to write-back. 
Each task begins with a bounded submission of ink (Section~\ref{sec:bounded-task-input}), selected canvas objects, and optional sigils; \tool then exposes its execution state (Section~\ref{sec:task-state}), returns consequential decisions to the user (Section~\ref{sec:consequential-decisions}), and writes reusable results back to the canvas (Section~\ref{sec:canvas-write-back}).

\subsubsection{Bounding Task Input through Ink and Canvas Objects}
\label{sec:bounded-task-input}

As illustrated in Figure~\ref{fig:bounded-task-input}, a bounded task input
combines ink expressions with user-selected canvas objects.
Users lasso the content they want to include and explicitly submit the
selection to the agent. The submitted ink may include both free-form ink and
optional handwritten sigils.

    \point{Ink expressions.}
    Every submission must include ink that expresses the user's intent within
    the canvas context. Ink expressions take two forms: unconstrained
    free-form ink and handwritten sigils with predefined semantics for
    explicitly specifying task resources or agent capabilities.

    \begin{enumerate}
        \item \textit{Free-form ink.}
        Free-form ink has no prescribed syntax. Users may write text, draw
        enclosures or arrows, connect objects, or combine these forms to
        express intent through both marks and spatial relationships.
        
        \item \textit{Handwritten sigils.}
        Handwritten sigils are compact ink expressions with predefined
        semantics that make particular task bindings explicit.
        A resource sigil \resourcesigil{} binds an existing canvas object to
        the task. After writing the sigil, users can bind an object by
        selecting it from the displayed list or directly on the canvas.
        A capability sigil \capabilitysigil{} binds the task to a named agent
        capability, implemented through a workflow, skill, or tool. Without
        an explicit capability binding, \tool infers an appropriate capability
        from the ink query and its context. When a capability is bound, it
        overrides this default inference and routes the task through the
        specified capability. The current prototype provides three common
        capabilities: finding resources, generating images, and organizing
        the canvas.
    \end{enumerate}

  \point{Canvas objects.}
  A submission may also include persistent canvas objects, such as text, images, PDFs, webpages, draw objects, and results from previous tasks, as shown on the left of
Figure~\ref{fig:system-architecture}.
  Users explicitly select or reference which objects to attach to the current ink query; these objects are submitted to the agent as task targets, supporting materials, or both. Because they retain their content, identity, source, and spatial position, users can provide them directly from the workspace without redescribing them in ink.

\subsubsection{Exposing and Controlling Task State}
\label{sec:task-state}

\begin{figure*}[t]
  \centering
  \includegraphics[width=\linewidth]{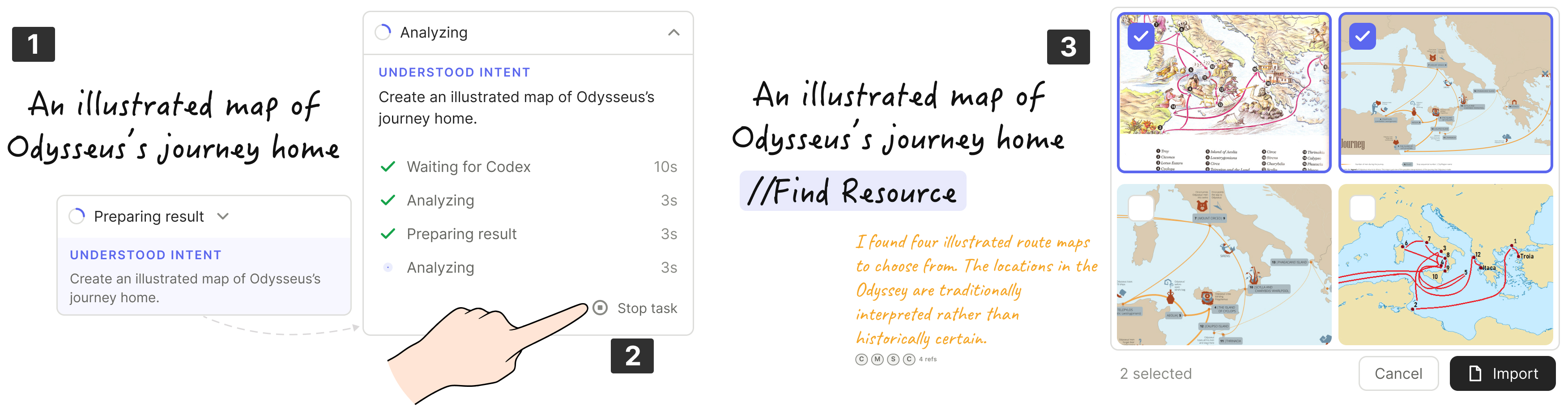}
  \caption{Inspecting and intervening in an agent task.
  (1) After the user submits a request for an illustrated map of Odysseus's journey, an agent activity card displays the understood intent and current activity at the location of the ink query.
  (2) Expanding the card reveals the activity history, current step, elapsed time, and a control for stopping the task.
  (3) When the resource search returns multiple candidates, the user selects which maps to import into the canvas.}
  \Description{A three-panel storyboard showing how a user inspects and intervenes in an agent task. In the first panel, an agent activity card displays the understood intent and current activity beside the submitted ink query. In the second panel, the expanded card shows completed and ongoing activities, elapsed time, and a stop control. In the third panel, four candidate maps are presented, two of which are selected for import.}
  \label{fig:task-observation-and-intervention}
\end{figure*}

\begin{figure*}[t]
  \centering
  \includegraphics[width=\linewidth]
  {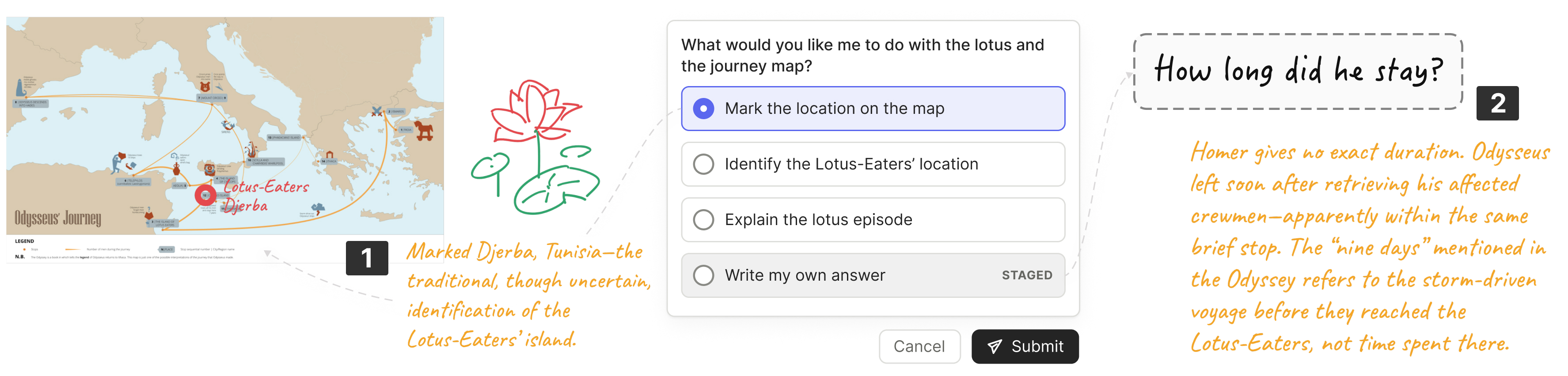}
  \caption{Resolving an ambiguous ink query through clarification.
  Amy draws a lotus beside an existing journey map, but the intended action is unclear, so \tool presents several possible interpretations and an option for a custom ink response.
  (1) Amy selects \textit{Mark the location on the map}, and \tool marks Djerba as the location traditionally associated with the Lotus-Eaters.
  (2) Alternatively, when the suggested options do not express her intent, Amy writes \textit{How long did he stay?}, which \tool interprets in the context of the lotus drawing and journey map.}
  \Description{A clarification interface and two possible outcomes.
  Amy has drawn a lotus beside a map of Odysseus's journey.
  The interface asks what she wants to do with the lotus and map and offers options to mark the location, identify the location, explain the episode, or write a custom response.
  In the first outcome, Djerba is marked on the map as the location traditionally associated with the Lotus-Eaters.
  In the second outcome, Amy writes a custom question asking how long Odysseus stayed, and the system provides a contextual answer.}
  \label{fig:task-clarification}
\end{figure*}

After submission, \tool creates an agent activity card associated with the originating location of the agent task.
The agent activity card progressively reveals information as the task proceeds.
Before an intent is available, it indicates that the request is being analyzed. 
During execution, its collapsed state shows the current activity; its expanded state additionally presents the understood intent, included canvas materials, completed and current activities, and elapsed time.
Users can stop a task that is no longer needed or is proceeding in the wrong direction, while an explicit failed state distinguishes execution failure from an active task.
This information represents the public task state needed for user judgment and intervention rather than the agent's private reasoning process.
By combining task interpretation, execution progress, and a stop control in one interface, \tool keeps an active agent action observable and interruptible.

Figure~\ref{fig:task-observation-and-intervention} illustrates this
mechanism through a request for an illustrated map of Odysseus's journey
home. 
The query was ambiguous and was interpreted as a request to generate a new
journey map, whereas the user intended to retrieve an existing, well-designed
map grounded in real geography from the web.
The user therefore stopped the task, added the capability sigil
\capabilitysigil{} to explicitly invoke resource retrieval, and
resubmitted the query.
The revised task returned a set of candidate maps for selection.

\subsubsection{Returning Consequential Decisions to the User}
\label{sec:consequential-decisions}

When missing information could materially affect the task direction, \tool does not continue from an uncertain interpretation.
Instead, it pauses the current action and requests clarification from the user.
When the system cannot determine how the submitted ink and canvas objects should be used together, the clarification interface presents a question and several possible interpretations at the relevant canvas location.
Users may select one of these options or provide a custom ink response.
The response remains staged so that it can be reviewed, cancelled, or
submitted before execution continues.

Figure~\ref{fig:task-clarification} illustrates two ways of resolving such ambiguity.
Amy vaguely recalls a lotus-related episode in \textit{The Odyssey} and draws a lotus beside the journey map, hoping to recover more information.
Because the drawing does not specify what she wants the agent to do, \tool asks how the lotus should be used with the map.
In the first path, Amy accepts one of the generated options and selects
\textit{Mark the location on the map}.
Following this clarification, \tool identifies Djerba as the location
traditionally associated with the Lotus-Eaters episode and marks it on the map (Figure~\ref{fig:task-clarification}~1).
In the second path, none of the generated options captures Amy's intent.
She therefore selects the custom-response option and writes \textit{``How long did he stay?''}.
Using the lotus drawing and journey map as context, \tool interprets \textit{``he''} as Odysseus and the question as asking about his stay during the Lotus-Eaters episode, then provides a corresponding answer (Figure~\ref{fig:task-clarification}~2).
By supporting both generated options and free-form ink responses, the clarification interface reduces the effort of common clarifications while preserving a way to express intentions that the system did not anticipate.

\begin{figure*}[t]
  \centering
  \includegraphics[width=0.99\textwidth]{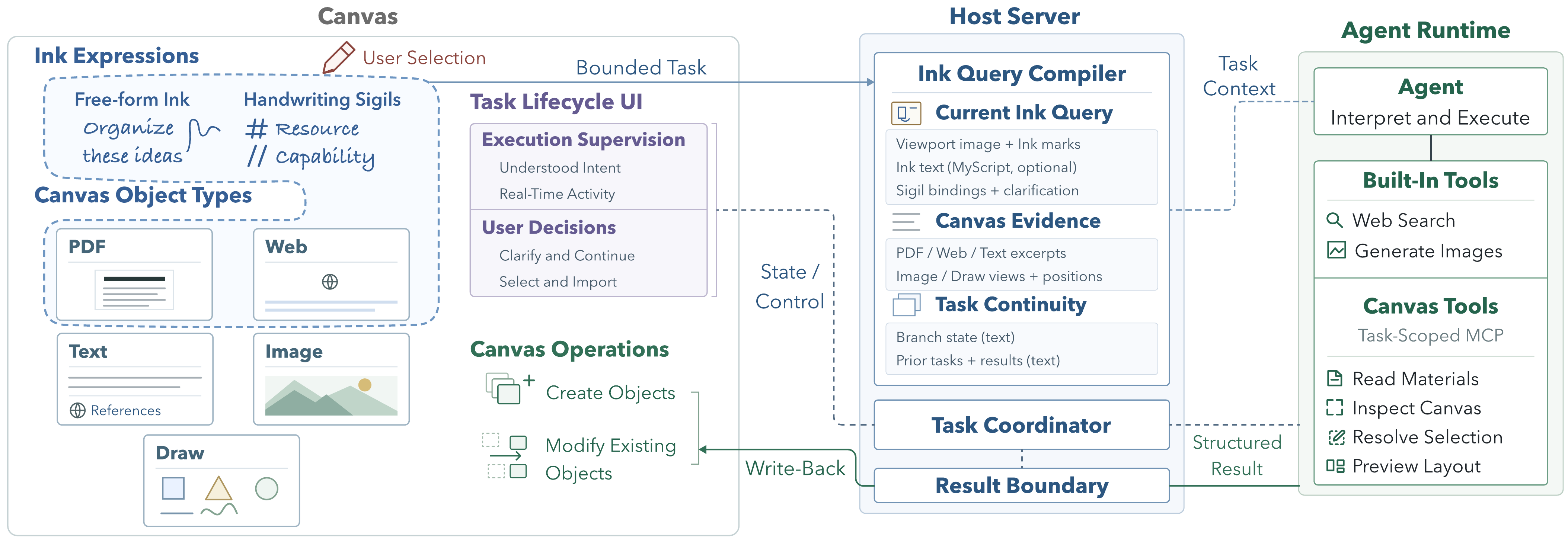}
  \caption{\tool's architecture for bounded agent task execution.
  Selected ink expressions, including free-form ink and handwritten sigils, are submitted with explicitly included canvas objects as a bounded task.
  The Ink Query Compiler constructs task context from the current ink query, canvas evidence, and task continuity.
  The Task Coordinator exchanges public state and user controls with the task lifecycle UI, while the agent runtime interprets and executes the task using built-in and task-scoped canvas tools.
  Structured results pass through the Result Boundary before Canvas Operations create or modify persistent objects.}
  \Description{A system architecture diagram organized into the persistent canvas, host server, and agent runtime.
  On the canvas, a user selection includes free-form ink, handwritten sigils, and selected canvas objects.
  The bounded task enters the host server, where the ink query compiler organizes the current ink query, canvas evidence, and task continuity into task context.
  A task coordinator exchanges execution state and user controls with the task lifecycle UI.
  The agent runtime interprets and executes the task using web search, image generation, and task-scoped tools for reading materials, inspecting the canvas, resolving selections, and previewing layouts.
  Structured results pass through a result boundary and return through canvas operations that create new objects or modify existing ones.}
  \label{fig:system-architecture}
\end{figure*}

\subsubsection{Writing Results Back to the Canvas}
\label{sec:canvas-write-back}

\tool writes structured results back through task-scoped canvas operations
that create new persistent objects or modify existing ones.
As shown on the left of Figure~\ref{fig:system-architecture}, write-back
determines the objects used to represent a result, how they enter the visual
structure of the canvas, and how their sources remain inspectable.
We organize this process around three aspects:

\point{Persistent and reusable objects.}
Through the \textit{Create Objects} operation shown in Figure~\ref{fig:system-architecture}, results become persistent canvas objects rather than remaining in a temporary task history.
Depending on their content, results may be represented as text, images, webpages, PDFs, or SVG-describable drawing primitives.
Users can subsequently move, edit, organize, and select these objects, and submit them with new ink as materials for later agent tasks.
They allow structures, relationships, and processes to be written back as editable visual results.

\point{Spatial arrangement.}
\tool uses spatial cues in the submitted ink to determine where and how results appear.
When no such cues are provided, it places results near the originating ink and relevant materials.
For semantically rich or logically structured results, \tool composes multiple canvas objects into a legible visual arrangement and uses \textit{Modify Existing Objects} to integrate the result with the canvas's existing visual structure.

\point{Embedded provenance.}
For generated text objects, \tool embeds compact source indicators directly within the object.
These indicators explicitly reference either external web sources or source-bearing resources registered on the canvas, such as webpage and PDF objects.
Users can inspect the sources supporting the generated text through these
indicators.
The source bindings remain attached as the text object is moved, organized,
or reused in a later task, rather than being separated into a global
reference list.

\subsection{Context Engineering and Agent Execution}

The preceding section described how users submit, supervise, and continue steering an agent task.
This section explains the system architecture that supports this lifecycle.
As shown in Figure~\ref{fig:system-architecture}, the Host Server mediates between the persistent canvas and a general-purpose agent runtime.
It constructs bounded task context, coordinates the task lifecycle, and controls how structured results are written back to the canvas.

\subsubsection{Constructing Bounded Task Context}

When the user submits a selection, the canvas sends its ink expressions, explicitly included canvas objects, and submission location to the host as a bounded task.
In addition to the explicitly submitted content, \tool uses the location of the ink query to define a bounded neighborhood of nearby ink and canvas objects.
These materials are registered as read-only context that the agent may inspect when needed.
Nearby content does not become a direct task input or operation target merely because of its proximity.
Explicit selection continues to determine which objects are included in and may be modified by the task.
The ink query compiler organizes the available information into three components: \textit{current ink query}, \textit{canvas evidence}, and \textit{task continuity}.
The current ink query retains a viewport image centered on the submission location, the original ink marks, optional MyScript transcription, resolved sigil bindings, and any clarification subsequently provided by the user.
Canvas evidence includes the explicitly submitted objects and provides on-demand access to nearby read-only materials.
PDF, webpage, and text objects contribute content excerpts, while image and drawing objects contribute visual representations and spatial positions.
Task continuity provides the state of the relevant work branch and textual representations of prior tasks and results, allowing a new ink query to continue an established line of work.
Together, these components form the task context sent to the agent runtime.
The agent can interpret the query using its explicit inputs, nearby canvas context, and relevant history without receiving unrestricted access to the entire canvas.

\subsubsection{Coordinating Agent Execution and Write-Back}

The agent runtime interprets and executes the task from the compiled task context.
In addition to built-in tools such as \textit{Web Search} and \textit{Generate Images}, it can use task-scoped canvas tools to read materials, inspect the canvas, resolve selections, and preview layouts.
These canvas tools allow the agent to gather task-relevant information and preview proposed arrangements, but they do not allow it to modify the persistent canvas directly.
During execution, the task coordinator sends public state, such as the understood intent and real-time activity, to the task lifecycle UI.
It also returns user controls, including stop requests and clarification responses, to the agent runtime.
When execution completes, the agent returns a structured result describing the objects, content, sources, and layout to be created or modified.
The structured result must pass through the result boundary before canvas operations can create new objects or modify existing objects within the task scope.
By separating general-purpose agent execution from persistent canvas mutation, this architecture bounds the context available to the agent, exposes its execution state, and controls the workspace scope that its results may change.

\subsection{Implementation}

\tool was implemented as a standalone web application optimized for iPad Safari.
The canvas client uses TypeScript, React~\cite{reactteam2026react}, and React Flow~\cite{xyflow2026reactflow} to implement the infinite canvas.
HTML Canvas 2D and Perfect Freehand~\cite{ruiz2026perfectfreehand} render pressure-sensitive vector ink, while the MyScript iink API~\cite{myscript2026iink} provides online handwriting transcription.
A Node.js and Fastify~\cite{fastifyteam2026fastify} host manages canvas state and agent tasks through HTTP.
The task coordinator streams public task state to the client through Server-Sent Events and relays user controls, including clarification responses and stop requests, to the agent runtime.
The host connects to an existing general-purpose Codex runtime through the Agent Client Protocol~\cite{zedindustries2026acp}.
The runtime can use built-in tools such as web search and image generation, together with task-scoped canvas tools provided through the Model Context Protocol~\cite{modelcontextprotocol2026mcp}.
These canvas tools allow the runtime to read materials, inspect the canvas, resolve selections, and preview proposed layouts without directly modifying the persistent workspace.
The study used GPT-5.6-Sol~\cite{openai2026gpt56sol} with low reasoning effort.

\point{Scenario-Based System Validation.}
Before the user study, we validated the end-to-end workflow of \tool using 34 distinct ink-query cases.
The validation examined the complete workflow from situated ink-query submission and context construction to agent execution and canvas write-back, rather than benchmarking model capability or evaluating user performance.
The cases covered different combinations of free-form ink, canvas object types, sigil bindings, and agent actions, including continuing tasks that reused results from earlier ink queries.
Stability did not require identical content across repeated generations.
Instead, every run had to produce a valid and semantically acceptable result while satisfying the expected canvas behavior.
Across the 34 cases, all 102 end-to-end runs completed successfully and met the stability requirements.

\section{User Study: Learning and Appropriating Ink-Based Agent Interaction}
\label{sec:user-study}

We examined how participants learn and use ink-initiated agent interaction in \tool through two research questions:

\begin{itemize}
    \item \textbf{RQ1:}
    How do participants learn to formulate ink queries and use them to complete increasingly complex knowledge-work tasks?

    \item \textbf{RQ2:}
    How do participants use ink to pursue their own goals, what activities and workflows emerge, and what possibilities and limitations does the system reveal?
\end{itemize}

To address these questions, we conducted two study phases.
In Reproduction, participants completed prescribed knowledge-work tasks of increasing complexity. In Freeform Creation, they pursued their own goals through open-ended exploration.

\subsection{Study Design}

\subsubsection{Participants and Apparatus}
Eighteen participants took part in the user study (12 female and 6 male; age $M_{age}=26.2$).
Participants included nine PhD students, seven master's or MPhil students, and two undergraduate students.
We recruited participants through public calls at a university and a research institute.
Eligible participants were required to have prior experience with either digital handwriting or visual canvas tools; remote participants were additionally required to provide their own stylus-enabled device.
Of the 18 participants, 10 participated in person and 8 participated remotely.
Both modes used the same browser-based version of \tool, study tasks, and materials.
In-person participants used researcher-provided device and stylus (iPad and Apple Pencil), while remote participants used their own stylus-enabled devices.
Sessions were scheduled for 60 minutes; when participants wished to continue during Freeform Creation, we extended that phase to allow them to explore the system more fully.
Each participant received compensation equivalent to approximately US\$9.
All participants provided informed consent before the study, including consent to screen and audio recording.

\subsubsection{Procedure}
The study session lasted approximately 1 hour per participant and consisted of an initial tutorial, Phase I (Reproduction), Phase II (Freeform Creation), and a concluding interview.

\paragraph{Introduction and Tutorial ($\sim$5 min).}
Before beginning the study tasks, participants entered a hands-on familiarization area.
They learned the basic canvas operations by using them, including writing, erasing, moving objects, and producing separate ink marks.
This stage familiarized participants with the input device and spatial workspace without demonstrating complete solutions to the subsequent study tasks.

\begin{figure*}[t]
  \centering
  \includegraphics[width=0.99\textwidth]{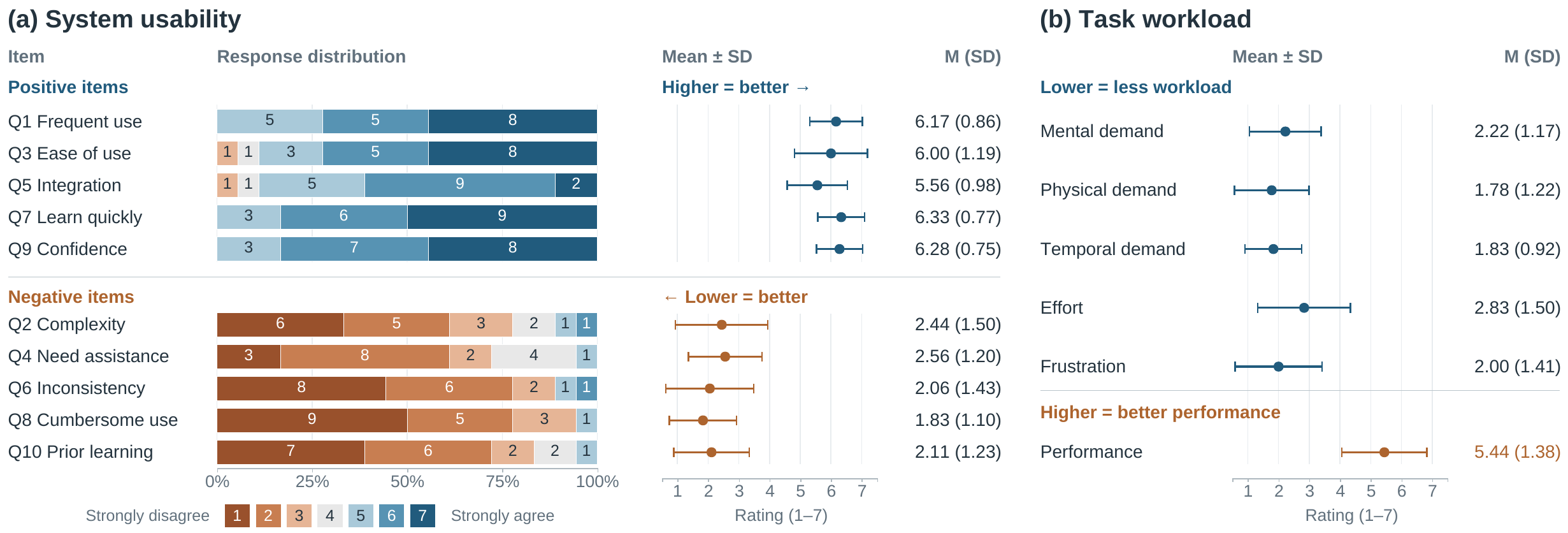}
  \caption{Post-study ratings from 18 participants on adapted seven-point scales: (a) SUS response distributions and item means; (b) NASA-TLX dimension means. Error bars indicate one standard deviation above and below the mean (M $\pm$ SD).}
  \Description{Two-panel chart of post-study questionnaire ratings. The SUS panel contains stacked response distributions and mean ratings for five positive and five negative items, separated by a horizontal line. Positive-item means range from 5.56 to 6.33, and negative-item means range from 1.83 to 2.56. The NASA-TLX panel shows means from 1.78 to 2.83 for five workload dimensions and a separate positively oriented performance mean of 5.44. All mean markers have error bars representing plus or minus one standard deviation.}
  \label{fig:quantitative-results}
\end{figure*}

\paragraph{Phase I: Reproduction ($\sim$25min).}
The Reproduction phase used a top-down, goal-directed design.
Participants worked through a shared progression of increasingly complex activities.
This phase examined whether they could understand the ink-query interaction and use it for continuing knowledge work.
The progression contained four levels:
\begin{enumerate}
  \item \textbf{L1. Steering with handwriting:} initiate an agent response and formulate a handwritten request for information, such as a local weather forecast.
  \item \textbf{L2.} \textbf{Bringing in useful materials \bringmaterials:} request different kinds of canvas content, including a generated image, research papers on a chosen topic, and the website of a conference, laboratory, or project.
  \item \textbf{L3.} \textbf{Making sense of materials \thinkwithmaterials:} ask a question grounded in an imported article or webpage, and identify a meaningful connection or difference across multiple materials.
  \item \textbf{L4.} \textbf{Turning materials into a structure \shapethespace:} organize retrieved papers into a visual account, such as a timeline including a comparison and summary. In our study, this final activity was presented as a bonus task.
\end{enumerate}
Rather than asking participants to reproduce a fixed final canvas, this phase asked them to progressively use a set of system capabilities under shared goals: direct ink submission, retrieval and generation, artifact-grounded ink queries, cross-material synthesis, and structured write-back.
Our analysis focuses on the completion of these interaction capabilities.

\paragraph{Phase II: Freeform Creation ($\sim$30min).}
The Freeform Creation phase used a bottom-up, open-ended design.
Participants started from a blank canvas and explored or created around a recent research direction or another topic of personal interest.
We prescribed neither a sequence of operations nor a uniform required output.
This phase examined the goals participants formulated, the activities and workflows they developed, the agent operations they attempted, and how they combined linguistic, spatial, relational, and visual ink to express their intentions.
It also allowed us to examine which self-initiated intentions \tool supported, which required clarification or repair, and which remained insufficiently supported, while collecting representative and unusual cases that reveal the interaction possibilities of ink queries.

\paragraph{Post-study measures and interview.}
After completing both phases, participants completed adapted 7-point versions of the System Usability Scale (SUS) and NASA Task Load Index (NASA-TLX) once for their overall experience with \tool, followed by a post-study interview.
Because these measures were administered after the complete session rather than separately after each phase, we interpret them as measures of overall perceived usability and workload and do not use them to claim a quantitative difference between Reproduction and Freeform Creation.

\subsubsection{Data Collection and Analysis}

Our data included screen and audio recordings, canvas interaction logs, ink-query and task histories, resulting canvas artifacts, questionnaires, and post-study interviews.
We analyzed these data in two ways:

\begin{enumerate}
    \item \textbf{Quantitative analysis.}
    We used descriptive statistics to summarize participants' overall SUS and NASA-TLX ratings and their interaction patterns. 
    For Reproduction, we coded task completion at four interaction levels (L1-L4) using interaction records and final canvas artifacts. 
    For Freeform Creation, we counted ink queries and summarized input-resource types, prior-output reuse, and other coded characteristics.

    \item \textbf{Qualitative analysis.}
    We qualitatively analyzed post-study interviews together with retained canvas artifacts and interaction records to understand participants' experiences, strategies, and perceived system boundaries.
    One author conducted the initial thematic coding, and a second author reviewed the codes and emerging themes; disagreements were resolved through discussion.
    We also examined sequences of related ink queries and their resulting artifacts to characterize evolving workflows and the compositional vocabulary through which participants combined linguistic, depictive, referential, relational, and spatial ink.
\end{enumerate}

\begin{figure*}[t]
  \centering
  \includegraphics[width=0.99\textwidth]{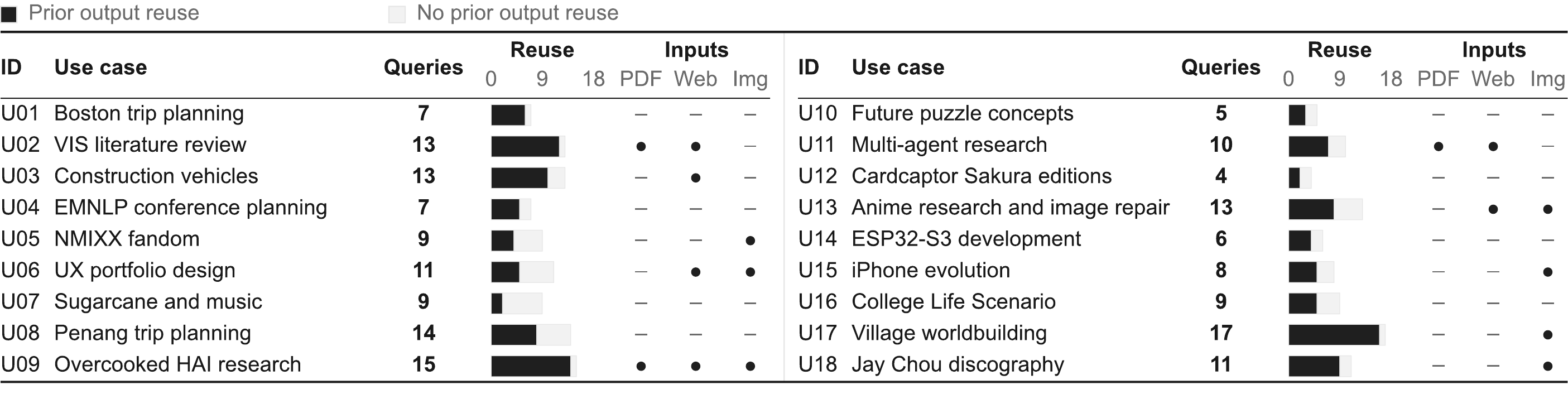}
  \caption{Overview of 18 Freeform Creation cases. Each row shows the use-case topic, total ink query count, counts with and without prior-output reuse, and the presence of PDF, web, or image inputs.}
  \Description{Two side-by-side tables list cases U01--U09 and U10--U18, with their activity topics and total ink query counts. Stacked bars distinguish prior output reuse and no prior output reuse. Presence markers indicate PDF, Web, and image inputs.}
  \label{fig:freeform-cases}
\end{figure*}

\subsection{Study Results}

We report participants' performance in Reproduction, the activities and patterns that emerged during Freeform Creation, and their feedback on the overall experience.

\subsubsection{Task Performance and Interaction Experience}

\paragraph{Reproduction Task Performance.}
\label{sec:reproduction-performance}
All participants learned to use ink to initiate and steer agent tasks on the canvas ({\small L1, 18/18}) and incorporated useful generated or retrieved materials into the canvas ({\small L2, 18/18}). 
Most participants also completed sensemaking activities such as explaining, summarizing, comparing, or revising existing materials ({\small L3, 16/18}). 
For the advanced or bonus visual-structuring task ({\small L4, 7/18}), 7 participants organized materials into temporal, conceptual, or workflow-based structures using spatial layout, connections, or annotations. 
Three partially completed the task, producing summaries or comparisons without forming a complete visual structure; the remaining eight did not complete it. 
Some participants did not continue in L3 and L4 because of time constraints or uncertainty about how to begin.

\paragraph{Perceived System Usability.}
All 18 participants completed both questionnaires without missing responses (Figure~\ref{fig:quantitative-results}).
The adapted 7-point SUS yielded an average score of 5.93 out of 7 ({\small $SD=0.25$}).
Standard SUS grade thresholds were not applied to this adapted scale. 
The highest-rated positive items concerned learnability ({\small $M=6.33$, $SD=0.77$}), confidence ({\small $M=6.28$, $SD=0.75$}), and willingness to use the system frequently ({\small $M=6.17$, $SD=0.86$}); all participants rated these items between 5 and 7. 
Negative-item ratings were low overall ({\small $M=1.83$--$2.56$}), although perceived need for assistance ({\small $M=2.56$, $SD=1.20$}) and unnecessary complexity ({\small $M=2.44$, $SD=1.50$}) received comparatively higher ratings among the negative items.

\paragraph{Perceived Task load.} 
Mean ratings on all five workload dimensions of the adapted NASA-TLX were below the scale midpoint: physical demand ({\small $M=1.78$, $SD=1.22$}), temporal demand ({\small $M=1.83$, $SD=0.92$}), frustration ({\small $M=2.00$, $SD=1.41$}), mental demand ({\small $M=2.22$, $SD=1.17$}), and effort ({\small $M=2.83$, $SD=1.50$}). The positively oriented performance item was favorable ({\small $M=5.44$, $SD=1.38$}), with 13/18 participants rating it between 5 and 7.

\subsubsection{Developing Work through Retained Ink and Materials}
\label{sec:open-ended-activities}
\label{sec:reflections-ink-reuse}

The Freeform Creation corpus comprised 18 cases and 181 ink query entries, including 14 clarification responses. Information-oriented activities were most prevalent ({\small $n=17$ cases}), followed by creative exploration ({\small $n=7$}) and speculative or playful exploration ({\small $n=2$}); these categories were not mutually exclusive. Across all 181 entries, more than two-thirds explicitly reused previously generated or retrieved materials for further questioning, explanation, comparison, or revision ({\small $n=125$, 69.1\%}), while the remaining 56 entries ({\small 30.9\%}) did not. Thirty entries combined multiple selected or linked canvas objects in a single request. These patterns show that prior outputs frequently became intermediate resources for ongoing activities rather than endpoints of individual requests. Figure~\ref{fig:freeform-cases} summarizes the activity topics and material reuse statistics across all cases.

Qualitative cases further showed how retained materials and ink supported continued work in three ways. First, in U16's college-life scenario, annotations of completed activities and changed circumstances informed the next year's plan; for later years, U16 wanted to modify and resubmit the existing request rather than rewrite it: \textit{``I do not want to copy it all over again...I can simply ask again.''} Second, retained instructions could be repurposed by changing their bindings or wording. We observed U12 reusing an instruction while rebinding it to a different canvas resource; in researcher-guided exchanges, U11 explored changing its target, revising individual words, and adding conditions such as \textit{short cards} before resubmission. Finally, participants considered how much instructional content should remain visible on the canvas. U17 preferred retaining handwritten keywords but requested speech for longer instructions to \textit{``both keep my canvas clean and let the AI know my intention''}; speech was a requested complement, not an evaluated feature. Together, these cases show that continued work relied on treating both prior outputs and their associated ink as editable resources while allowing users to manage how much content remained on the canvas.

\subsubsection{Referring to Fine-Grained Targets within Canvas Objects}
\label{sec:reflections-sigils-local-reference}

Participants used ink to refer to targets within canvas objects at a finer granularity than the interface's explicit bindings. While selection and resource-sigil binding operated at the level of whole objects, participants marked passages or image details within them, including U01's reference to one day in an itinerary, U13's marks on particular characters in a generated group, and U17's local map annotations (Sections~\ref{sec:ink-vocabulary} and~\ref{sec:gallery-episodes}). In the current implementation, the agent handled these references primarily by interpreting free-form ink together with object content, without dedicated interaction support for specifying or confirming their fine-grained scope. This support was not always reliable: U07 asked about the song \textit{Sugar} within a text object, but the response covered all eight songs, while U17's local map edits required repeated revisions. The observed outcomes do not reveal whether these mismatches arose during reference interpretation or action execution. Together, these cases show that within-object reference had already emerged as a meaningful use of ink, while supporting its specification, inspection, and preservation more reliably remains an important direction for system improvement.

\begin{figure*}[t]
  \centering
  \includegraphics[width=0.99\textwidth]{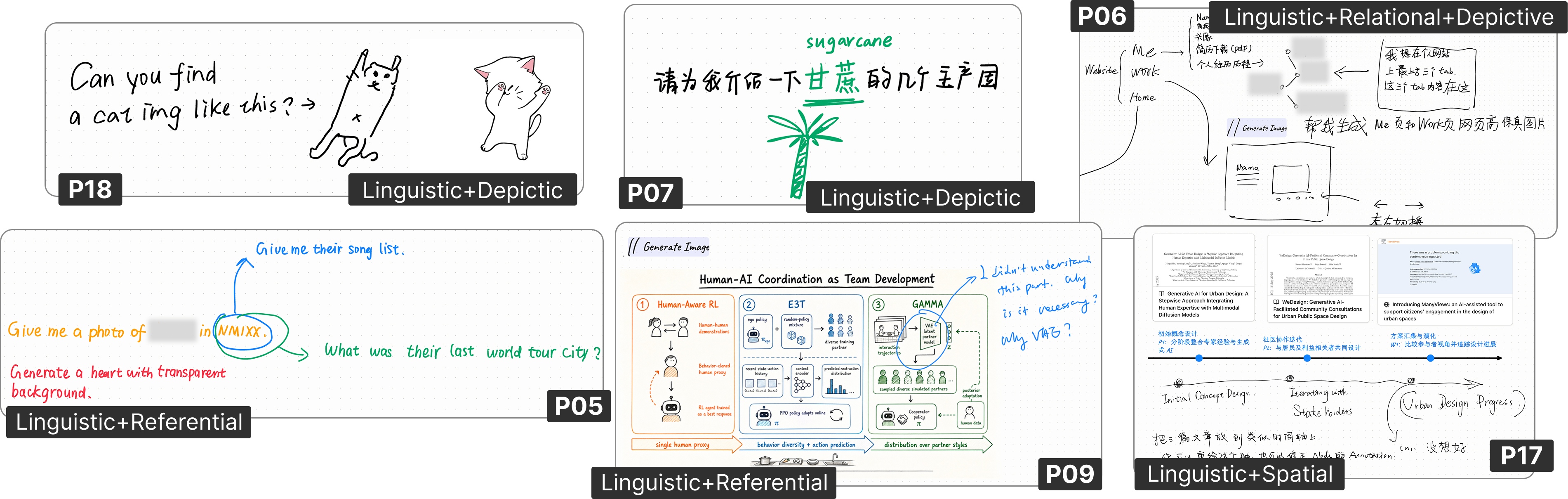}
  \caption{Compositional roles of ink in Freeform Creation. Participants typically combined a linguistic request with depictive, referential, relational, or spatial ink, rather than using these forms independently.}
  \Description{Five rows pair original ink excerpts with examples and counts. Linguistic shows a handwritten Boston travel request. Depictive shows a Chinese question about sugarcane-producing countries with a plant drawing and the English label sugarcane. Referential / Scope-marking shows a circle around Day 2 of an itinerary beside a revision request. Relational shows a branching website diagram with Me, Work, and Home. Spatial / Layout shows a curved line with marked endpoints and the handwritten instruction Using the line.}
  \label{fig:ink-vocabulary}
\end{figure*}

\subsubsection{Seeking Outputs That Support Further Work}
\label{sec:reflections-output-materials}

Participants evaluated outputs by how they could read, organize, and revisit them on the canvas, not only by the information they contained. Three expectations emerged. First, several participants preferred visually structured responses to a single or densely written text object. U14 requested larger headings, smaller descriptions, and branching levels; U08 valued structured presentation, and U09 associated visualization with a more concise and convenient canvas experience. This did not imply a uniform preference for less text, as U07 considered the explanations received appropriately detailed. Second, participants distinguished an arrangement of constituent elements from a single image depicting a similar structure. U14 asked for a flowchart made with \textit{``pen strokes and arrows, rather than an image''}, while U15 wanted \textit{``a whole set of images, then arranged into a timeline''} of iPhone technology but received one image instead. Finally, U17 wanted to retain different versions together with \textit{``what I changed at each step, what my goal was, and what the generated image looked like, all on the canvas''}. Together, these responses indicate that participants wanted outputs whose visual structure, constituent elements, and evolution remained available for inspection and revision during subsequent work.

\section{Case Gallery: Ink-Initiated Agent Interaction in Practice}
\label{sec:case-gallery}

\begin{figure*}[t]
  \centering
  \includegraphics[width=0.99\textwidth]{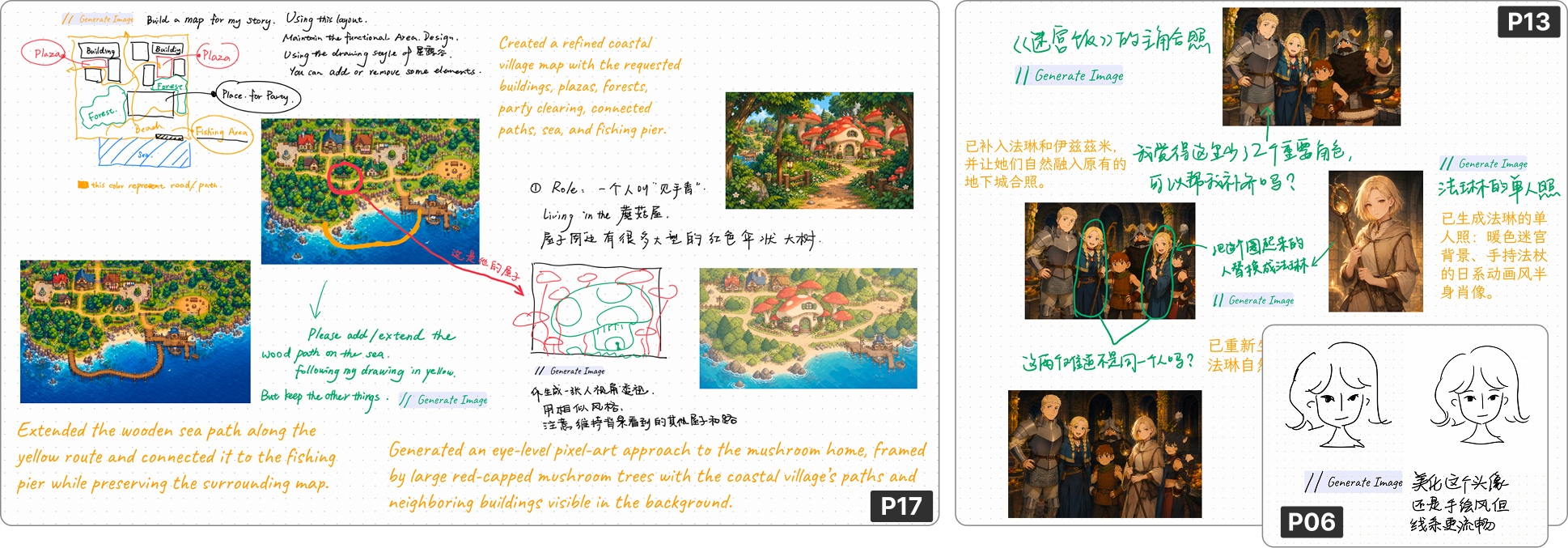}
  \caption{Three trajectories of ink-guided image creation during Freeform Creation. P17 developed a coastal-village concept from a spatial sketch and locally annotated generated maps and scenes; P13 used circles and handwritten instructions to guide successive character revisions; and P06 used a hand-drawn avatar as a reference for stylistic refinement. Across the cases, participants combined linguistic requests with sketches, references, and local annotations to iteratively develop image content, composition, and style.}
  \Description{A collage of three image-creation cases. The large panel on the left shows P17's labeled village map, generated map variants, annotations that revise a route to the sea, and a mushroom-house sketch and generated scene. The upper-right panel shows P13's successive character-group images with handwritten instructions and circles around selected characters. The lower-right inset shows P06's hand-drawn avatar and a refined version guided by a request to retain its hand-drawn line style.}
  \label{fig:user-case-ink-image}
\end{figure*}

\begin{figure*}[t]
  \centering
  \includegraphics[width=0.99\textwidth]{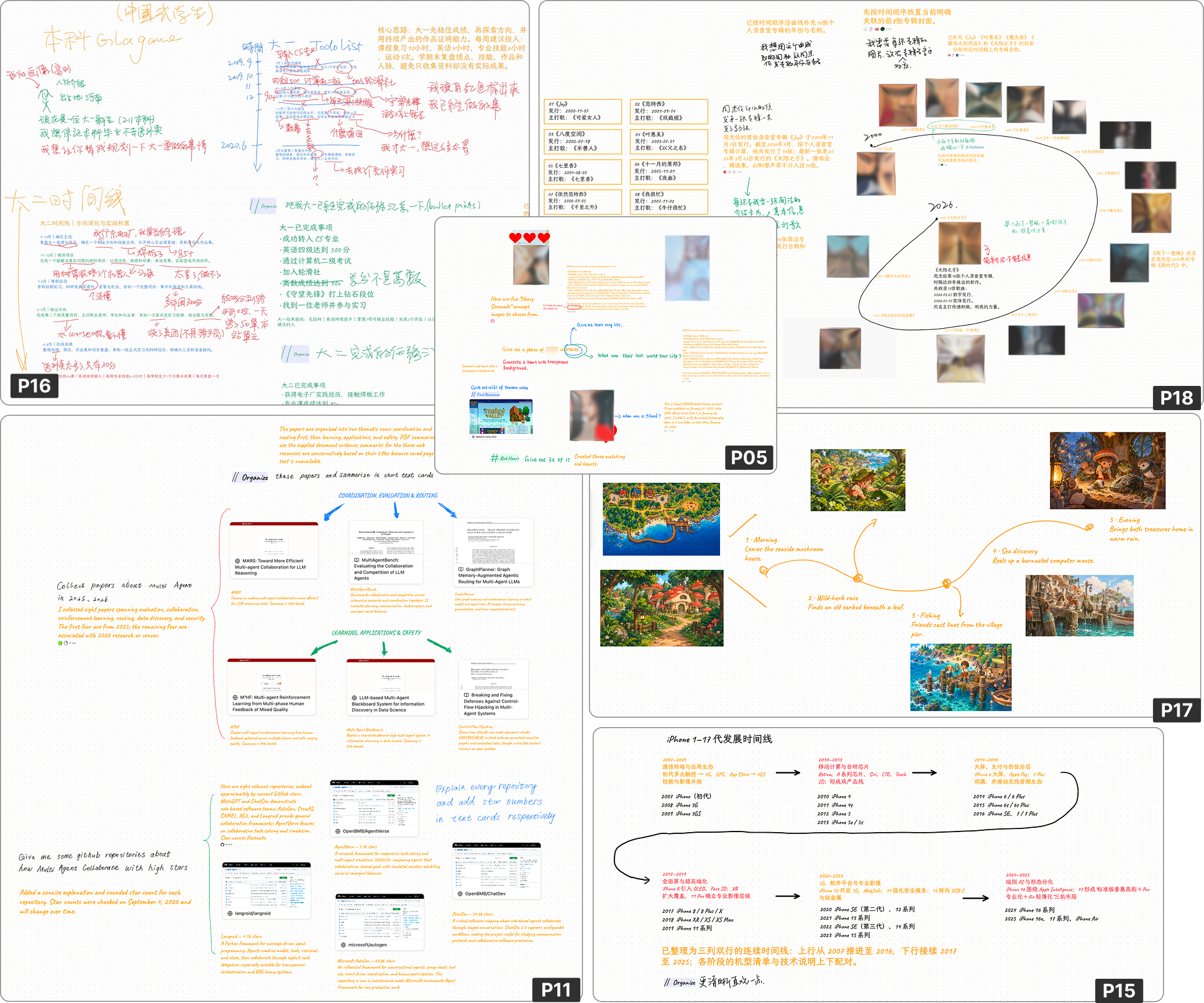}
  \caption{Six examples of visual structures developed through ink queries during Freeform Creation. Participants organized collected or generated materials as annotated timelines (P16, P18, and P15), thematic branches and card groups (P11), a topic-centered constellation (P05), and a spatial narrative sequence (P17). These arrangements made temporal, semantic, and narrative relationships among materials visible and available for continued work.}
  \Description{A collage of six structured canvases. P16 combines generated plans and multicolored annotations in staged timelines. P18 places album covers and information along a curved chronology. P05 arranges images, notes, and requests around a shared music-related topic. P11 groups research-paper cards into two thematic branches and places related software repositories below them. P17 connects generated village scenes along an orange narrative path. P15 organizes successive iPhone generations along a continuous two-row timeline.}
  \label{fig:user-case-ink-space}
\end{figure*}

This gallery examines how participants composed ink queries and used them to guide agent actions during Freeform Creation. Complementing the aggregate results in Section~\ref{sec:user-study}, we first characterize this compositional vocabulary, then trace how it supported continuing work across selected episodes of design, scenario exploration, and storytelling.

\subsection{Ink Query Vocabulary}
\label{sec:ink-vocabulary}

What did users do with ink when they could choose their own goals? They did not simply replace words with drawings. Often, they wrote the question or action, then used a circle, a sketch, a connection, or a path to show the rest: which part they meant, what something should look like, how ideas belonged together, or where an answer should go.
Percentages indicate observed occurrences across all 181 ink query entries; a single entry could involve multiple uses of ink.
Figure~\ref{fig:ink-vocabulary} pairs five submitted-ink excerpts with the small stories behind them, illustrating five uses of ink:
\begin{itemize}
    \item \textbf{Linguistic (97.8\%):} Handwritten words, micro-prompts~\cite{gmeiner2025intent}, or sentences convey requests or information.
    \item \textbf{Depictive (9.9\%):} Sketches depict visual forms or content relevant to the request.
    \item \textbf{Referential / scope-marking (39.2\%):} Marks point to or enclose existing content to identify the target or scope of a request, referring either to entire canvas objects or to specific details within them, such as words or sentences in a text object or regions of an image.
    \item \textbf{Relational (6.1\%):} Connecting marks express relationships between elements, providing a visual structure that reflects the logical organization of content.
    \item \textbf{Spatial / layout (9.4\%):} Ink placement indicates the intended spatial organization of content.
\end{itemize}

These uses of ink often appeared together. 
In 76 of 181 ink query entries, ink served at least two confirmed functions; among the 150 records where all five functions could be assessed, writing combined with marks pointing to a target was the most common multi-function combination ({\small $n=46$}). The itinerary example captures this well: the sentence says what should change, while the circle says where to look. Drawings, relations, and layout cues appeared in fewer cases, but brought other possibilities into view, especially in design and storytelling.
The interesting vocabulary is therefore not a dictionary in which every circle or arrow has one meaning. An arrow can point to a detail, connect two ideas, or direct an answer toward a space. For future systems, the practical lesson is to keep these cues together: read the words, but also follow what the user has circled, drawn, connected, or laid out.

\subsection{Selected Episodes of Open-Ended Use}
\label{sec:gallery-episodes}

\subsubsection{Exploring image creation through ink.}
Figure~\ref{fig:user-case-ink-image} highlights three selected episodes in which participants used ink to progressively develop images. P17 first drew a labeled spatial sketch of a coastal village, then annotated generated maps to extend a route toward the sea, and finally sketched a mushroom house to create a new scene within the same setting. P13 treated successive character images as intermediate results, circling particular characters and adding handwritten instructions to guide local changes in their appearance and grouping. P06 instead used a hand-drawn avatar as a visual reference and asked the agent to refine it while preserving its hand-drawn line style. These episodes illustrate how sketches, linguistic specifications, and referential annotations could work together across multiple ink queries to develop image content, composition, and style.

\subsubsection{Creating visually structured outcomes.}
Figure~\ref{fig:user-case-ink-space} presents selected cases in which participants transformed collected or generated information into distinctive visual structures. Timeline structures appeared in P16's staged scenario, P18's chronological arrangement of albums, and P15's history of successive product generations, while P17 connected generated scenes along a path to form a visual narrative. P11 organized research papers as card-like objects and connected thematic groups through a tree-shaped branching structure, making both individual materials and their conceptual relationships visible. P05 developed a more radial, star-like structure by placing a shared topic at the center and arranging related images, information, and follow-up requests around it. Together, these cases illustrate how participants used temporal ordering, cards, branches, and radial arrangements to give collected materials a visible organizational logic on the canvas.

\section{Discussion}
\label{sec:discussion}

Drawing on the study findings, we discuss the value of ink for directing agents and the challenges of supporting continued work through it.

\subsection{The Distinctive Value of Ink for Steering Agents}

Our study highlights ink's value as a way to express requests directly through the materials and visual structures with which users are working.
Participants combined brief handwritten instructions with depictions, references to materials, and marks expressing relationships or layout (Section~\ref{sec:ink-vocabulary}).
Words, marks, and relevant materials thus jointly expressed the request.
This connects handwriting and sketching as resources for developing ideas~\cite{riche2017aswemayink,kim2019inking} with the joint construction of instructions through language, interaction, and materials~\cite{shen2026iai}.

Within this combination, ink can both refer to what is already present and express what is not yet there.
Selection and drag-and-drop also ground requests in existing content~\cite{masson2024directgpt,wang2025vizta}, but referring to an existing object is only part of what our cases illustrate.
A sketch can depict a proposed form, while connections or a path can specify relationships or an arrangement to create.
These marks let users show aspects of the intended result alongside the materials it should build on, without first describing every visual or spatial detail in words.
Selected materials and prior results can in turn supply what a brief expression leaves implicit.
Work on contextual sketch interpretation similarly shows how marks acquire meaning in relation to the material being worked on~\cite{yen2025codeshaping}.
Ink thus makes graphical and spatial expression available alongside language as part of the request.

Ink also lets users develop the content of a request as they work through an idea.
Annotations could develop the content on which a request depended, while an earlier instruction could be revised or applied to another material.
The retained-ink findings (Section~\ref{sec:reflections-ink-reuse}) and the design and storytelling episodes (Section~\ref{sec:gallery-episodes}) show how composing the work and composing a request can overlap.
Building on actionable annotations and iterative ink-based ideation~\cite{romat2019activeink,payandeh2026thinkink}, our study illustrates how this overlap can operate across acquiring, examining, and organizing materials.
The effort invested in an annotation or sketch can therefore contribute both to developing the user's ideas and to expressing the assistance they seek, with retained ink remaining available for further requests.

This value can coexist with complementary input for longer explanations.
The preference for retaining useful annotations and keywords while supplementing instructions with speech (Section~\ref{sec:reflections-ink-reuse}) suggests a way to combine ink's graphical and spatial expression with spoken or typed detail, consistent with sketch-and-speech interaction~\cite{huang2025sketchgpt}.
Ink is worth supporting because users can seek assistance through the same handwriting and sketching with which they formulate, inspect, and revise their ideas.

\subsection{Supporting Continued Work in Ink-Initiated Agent Interfaces}

Connecting material acquisition, inquiry, and organization allows agent assistance to participate in work whose next steps develop through engagement with earlier results.
Participants reused outputs to ask further questions, add changed circumstances, and revise emerging content (Section~\ref{sec:open-ended-activities}).
These practices accord with sensemaking accounts in which people gather information and revise interpretations through external representations~\cite{russell1993cost,pirolli2005sensemaking,andrews2010spacetothink}.
In \tool, ink provides a way to re-enter this process at the material under consideration: users can annotate a result, connect it to other content, or sketch how it should develop, then seek assistance with that next step.
Building on actionable ink and iterative ideation~\cite{romat2019activeink,payandeh2026thinkink}, this suggests embedding agent interaction within the ongoing handling of materials, so that users can develop what to ask through examining and working on what they receive.

For this participation to be useful, results need to fit the ways users read, organize, and develop their work.
Participants' preferences for visual hierarchy and constituent elements (Section~\ref{sec:reflections-output-materials}) concerned how an answer entered the workspace, as well as its content.
This complements work on reusable conversational materials and spatially organized writing~\cite{yen2024memolet,choe2026crosslit}: the representation of a response matters to both understanding it and working with it afterward.
Designing outputs \emph{for} the canvas therefore means considering how content and relationships can remain readable and available for further annotation or revision.
A structure assembled from separate elements and an image depicting that structure offer different ways to continue; either form needs interaction support suited to the material.
The local-reference findings (Section~\ref{sec:reflections-sigils-local-reference}) make this need concrete, since users could mark details within an object even when the response did not maintain that local scope.
Fine-grained reference mechanisms~\cite{wang2025vizta} offer one route for connecting these marks to the relevant passages or regions.

Users also need to retain direction over how agent assistance enters an evolving activity.
When a new annotation draws on existing materials and prior results, its meaning depends on which context the agent uses and which changes the user intends.
Making the interpreted task available for inspection and correction could let users adjust that meaning as their work develops, following research on interaction-augmented instructions and correctable AI behavior~\cite{shen2026iai,amershi2019guidelines}.
In \tool, task inspection and clarification provide initial support for this involvement.
The broader opportunity is to keep users' judgments about the current materials connected to the agent's next action, including when users revise their goals or find that a result calls for a different direction.

This involvement can extend from guiding the next action to reviewing the development of the work.
U17's expectation of retaining versions alongside modification intentions (Section~\ref{sec:reflections-output-materials}) suggests keeping changes and their purposes available together, so users can compare outputs in relation to what they were trying to achieve.
For other applications built around reading, annotation, or visual organization, the transferable direction is to let users seek and guide assistance through the materials they are already developing.
Ink-based interaction could then accompany the movement between examining a result, revising an idea, and deciding what assistance to seek next.

\subsection{Limitations and Future Work}

\paragraph{Study setting and real-world use.}
The study was limited to a single session with 18 undergraduate and graduate students who had prior experience with digital handwriting or visual canvas tools.
How these practices extend to other user populations and sustained everyday work remains to be explored.
The evaluation was descriptive rather than comparative, leaving the relative benefits of ink-based interaction over alternative approaches open for further study.
Longitudinal field deployments with a broader range of users could explore how ink-based agent interaction fits into real-world work, how personal ink vocabularies and continuing workflows develop, and when users choose ink alongside other forms of interaction.

\paragraph{Interpretation and personalization.}
The expressive flexibility of free-form ink also introduces uncertainty in how an agent interprets a request.
Rather than relying on predefined mappings between input and action, \tool depends on the model's interpretation of ink in context, which may diverge from the user's intended meaning.
Advances in contextual ink understanding(e.g., online handwriting recognition, and vision models) could strengthen the reliability of this interaction.
Because users differ in how they express ideas and requests through ink, future work could also explore personalized learning from users' corrections and repeated use to adapt interpretation to their individual conventions and preferences.

\section{Conclusion}
\label{sec:conclusion}

This work explored how ink can enable users to seek and steer agent assistance in ongoing knowledge work.
A formative study identified three connected activities and interaction expectations that informed the design of \tool.
The system supports bringing materials in, thinking with materials, and shaping the space, with acquired materials and subsequent results retained in the same workspace for further work.
A user study with 18 participants found favorable perceived usability and low workload, while highlighting how retained ink and materials supported further inquiries and revealing challenges with local references and output form.
An ink-query vocabulary characterized intention expression, while cases illustrated ink's potential for guiding creation, iterative revision, and visual organization.
These findings highlight ink's potential to make agent assistance part of how people think and work with materials, drawing on the language, sketches, and spatial marks through which they express and develop ideas.


\bibliographystyle{ACM-Reference-Format}
\bibliography{reference}

@inproceedings{riche2017aswemayink,
author = {Riche, Yann and Henry Riche, Nathalie and Hinckley, Ken and Panabaker, Sheri and Fuelling, Sarah and Williams, Sarah},
title = {As We May Ink? Learning from Everyday Analog Pen Use to Improve Digital Ink Experiences},
year = {2017},
isbn = {9781450346559},
publisher = {Association for Computing Machinery},
address = {New York, NY, USA},
url = {https://doi.org/10.1145/3025453.3025716},
doi = {10.1145/3025453.3025716},
booktitle = {Proceedings of the 2017 CHI Conference on Human Factors in Computing Systems},
pages = {3241–3253},
numpages = {13},
location = {Denver, Colorado, USA},
series = {CHI '17}
}

@inproceedings{kidd1994marks,
  author = {Kidd, Alison},
  title = {The marks are on the knowledge worker},
  year = {1994},
  isbn = {0897916506},
  publisher = {Association for Computing Machinery},
  address = {New York, NY, USA},
  url = {https://doi.org/10.1145/191666.191740},
  doi = {10.1145/191666.191740},
  booktitle = {Proceedings of the SIGCHI Conference on Human Factors in Computing Systems},
  pages = {186–191},
  numpages = {6},
  location = {Boston, Massachusetts, USA},
  series = {CHI '94}
}

@inproceedings{intenttagging2025gmeiner,
author = {Gmeiner, Frederic and Marquardt, Nicolai and Bentley, Michael and Romat, Hugo and Pahud, Michel and Brown, David and Roseway, Asta and Martelaro, Nikolas and Holstein, Kenneth and Hinckley, Ken and Riche, Nathalie},
title = {Intent Tagging: Exploring Micro-Prompting Interactions for Supporting Granular Human-GenAI Co-Creation Workflows},
year = {2025},
isbn = {9798400713941},
publisher = {Association for Computing Machinery},
address = {New York, NY, USA},
url = {https://doi.org/10.1145/3706598.3713861},
doi = {10.1145/3706598.3713861},
booktitle = {Proceedings of the 2025 CHI Conference on Human Factors in Computing Systems},
articleno = {531},
numpages = {31},
location = {Yokohama, Japan},
series = {CHI '25}
}

@inproceedings{lin2025inkspire,
author = {Lin, David Chuan-En and Kang, Hyeonsu B. and Martelaro, Nikolas and Kittur, Aniket and Chen, Yan-Ying and Hong, Matthew K.},
title = {Inkspire: Supporting Design Exploration with Generative AI through Analogical Sketching},
year = {2025},
isbn = {9798400713941},
publisher = {Association for Computing Machinery},
address = {New York, NY, USA},
doi = {10.1145/3706598.3713397},
booktitle = {Proceedings of the 2025 CHI Conference on Human Factors in Computing Systems},
articleno = {427},
numpages = {18},
location = {Yokohama, Japan},
series = {CHI '25}
}

@misc{tldraw2026,
  author = {{tldraw}},
  title = {tldraw: A collaborative digital whiteboard},
  year = {2026},
  howpublished = {\url{https://tldraw.com/}},
  note = {Accessed: 2026-10-29}
}

@inproceedings{shi2026notational,
author = {Shi, Xinyu and Wei, Li-Yi and Zhao, Nanxuan and Zhao, Jian and Kazi, Rubaiat Habib},
title = {Notational Animating: An Interactive Approach to Creating and Editing Animation Keyframes},
year = {2026},
isbn = {9798400722783},
publisher = {Association for Computing Machinery},
address = {New York, NY, USA},
doi = {10.1145/3772318.3790707},
booktitle = {Proceedings of the 2026 CHI Conference on Human Factors in Computing Systems},
articleno = {814},
numpages = {27},
location = {Barcelona, Spain},
series = {CHI '26}
}

@inproceedings{leung2026annotategpt,
author = {Leung, Benedict and Shimabukuro, Mariana and Collins, Christopher},
title = {AnnotateGPT: Designing Human–AI Collaboration in Pen-Based Document Annotation},
year = {2026},
isbn = {9798400722783},
publisher = {Association for Computing Machinery},
address = {New York, NY, USA},
url = {https://doi.org/10.1145/3772318.3790867},
doi = {10.1145/3772318.3790867},
booktitle = {Proceedings of the 2026 CHI Conference on Human Factors in Computing Systems},
articleno = {811},
numpages = {25},
location = {
},
series = {CHI '26}
}

@inproceedings{gmeiner2025intent,
  author = {Gmeiner, Frederic and Marquardt, Nicolai and Bentley, Michael and Romat, Hugo and Pahud, Michel and Brown, David and Roseway, Asta and Martelaro, Nikolas and Holstein, Kenneth and Hinckley, Ken and Henry Riche, Nathalie},
  title = {Intent Tagging: Exploring Micro-Prompting Interactions for Supporting Granular Human-{GenAI} Co-Creation Workflows},
  booktitle = {Proceedings of the 2025 CHI Conference on Human Factors in Computing Systems},
  series = {CHI '25},
  year = {2025},
  pages = {1--31},
  numpages = {31},
  publisher = {Association for Computing Machinery},
  address = {New York, NY, USA},
  doi = {10.1145/3706598.3713861},
  url = {https://doi.org/10.1145/3706598.3713861}
}

@inproceedings{marquardt2025imaginationvellum,
author = {Marquardt, Nicolai and Roseway, Asta and Romat, Hugo and Panda, Payod and Pahud, Michel and Ramos, Gonzalo and Drucker, Steven M. and Wilson, Andrew D. and Hinckley, Ken and Riche, Nathalie},
title = {ImaginationVellum: Generative-AI Ideation Canvas with Spatial Prompts, Generative Strokes, and Ideation History},
year = {2025},
isbn = {9798400720376},
publisher = {Association for Computing Machinery},
address = {New York, NY, USA},
url = {https://doi.org/10.1145/3746059.3747631},
doi = {10.1145/3746059.3747631},
booktitle = {Proceedings of the 38th Annual ACM Symposium on User Interface Software and Technology},
articleno = {159},
numpages = {19},
location = {
},
series = {UIST '25}
}

@inproceedings{romat2019activeink,
  author    = {Hugo Romat and Nathalie Henry Riche and Ken Hinckley and Bongshin Lee and Caroline Appert and Emmanuel Pietriga and Christopher Collins},
  title     = {{ActiveInk}: (Th)Inking with Data},
  booktitle = {Proceedings of the 2019 CHI Conference on Human Factors in Computing Systems},
  series    = {CHI '19},
  year      = {2019},
  pages     = {1--13},
  publisher = {ACM},
  doi       = {10.1145/3290605.3300272},
}

@inproceedings{kim2019inking,
  author    = {Yea-Seul Kim and Nathalie Henry Riche and Bongshin Lee and Matthew Brehmer and Michel Pahud and Ken Hinckley and Jessica Hullman},
  title     = {Inking Your Insights: Investigating Digital Externalization Behaviors During Data Analysis},
  booktitle = {Proceedings of the 2019 ACM International Conference on Interactive Surfaces and Spaces},
  series    = {ISS '19},
  year      = {2019},
  pages     = {255--267},
  publisher = {ACM},
  doi       = {10.1145/3343055.3359714},
}

@inproceedings{subramonyam2020texsketch,
  author    = {Hariharan Subramonyam and Colleen M. Seifert and Priti Shah and Eytan Adar},
  title     = {{texSketch}: Active Diagramming through Pen-and-Ink Annotations},
  booktitle = {Proceedings of the 2020 CHI Conference on Human Factors in Computing Systems},
  series    = {CHI '20},
  year      = {2020},
  pages     = {1--13},
  publisher = {ACM},
  doi       = {10.1145/3313831.3376155},
}

@inproceedings{shi2024inkeraction,
  author    = {Lei Shi and Rachel Campbell and Peggy Chi and Maria Cirimele and Mike Cleron and Kirsten Climer and Chelsey Q Fleming and Ashwin Ganti and Philippe Gervais and Pedro Gonnet and Tayeb A Karim and Andrii Maksai and Chris Melancon and Rob Mickle and Claudiu Musat and Palash Nandy and Xiaoyu Iris Qu and David Robishaw and Angad Singh and Mathangi Venkatesan},
  title     = {{Inkeraction}: An Interaction Modality Powered by Ink Recognition and Synthesis},
  booktitle = {Proceedings of the CHI Conference on Human Factors in Computing Systems},
  series    = {CHI '24},
  year      = {2024},
  pages     = {1--26},
  publisher = {Association for Computing Machinery},
  address   = {New York, NY, USA},
  doi       = {10.1145/3613904.3642498},
  url       = {https://doi.org/10.1145/3613904.3642498}
}

@article{lin2024inksight,
  author    = {Yanna Lin and Haotian Li and Leni Yang and Aoyu Wu and Huamin Qu},
  title     = {{InkSight}: Leveraging Sketch Interaction for Documenting Chart Findings in Computational Notebooks},
  journal   = {IEEE Transactions on Visualization and Computer Graphics},
  volume    = {30},
  number    = {1},
  year      = {2024},
  pages     = {944--954},
  doi       = {10.1109/TVCG.2023.3327170},
}

@inproceedings{yen2025codeshaping,
  author    = {Ryan Yen and Jian Zhao and Daniel Vogel},
  title     = {Code Shaping: Iterative Code Editing with Free-form AI-Interpreted Sketching},
  booktitle = {Proceedings of the 2025 CHI Conference on Human Factors in Computing Systems},
  series    = {CHI '25},
  year      = {2025},
  articleno = {872},
  numpages  = {17},
  publisher = {ACM},
  doi       = {10.1145/3706598.3713822},
}

@inproceedings{li2026sketchdynamics,
  author    = {Boyu Li and Lin-Ping Yuan and Zeyu Wang and Hongbo Fu},
  title     = {{SketchDynamics}: Exploring Free-Form Sketches for Dynamic Intent Expression in Animation Generation},
  booktitle = {Proceedings of the 2026 CHI Conference on Human Factors in Computing Systems},
  series    = {CHI '26},
  year      = {2026},
  articleno = {886},
  numpages  = {21},
  publisher = {ACM},
  doi       = {10.1145/3772318.3791071},
}

@inproceedings{huang2025sketchgpt,
  author    = {Zeyuan Huang and Cangjun Gao and Yaxian Shan and Haoxiang Hu and Qingkun Li and Xiaoming Deng and Cuixia Ma and Yu-Kun Lai and Yong-Jin Liu and Feng Tian and Guozhong Dai and Hongan Wang},
  title     = {{SketchGPT}: A Sketch-based Multimodal Interface for Application-Agnostic LLM Interaction},
  booktitle = {Proceedings of the 38th Annual ACM Symposium on User Interface Software and Technology},
  series    = {UIST '25},
  year      = {2025},
  articleno = {157},
  numpages  = {18},
  publisher = {ACM},
  doi       = {10.1145/3746059.3747598},
}

@inproceedings{ma2023sketchingrelatedwork,
  author    = {Donghyeok Ma and Joon Hyub Lee and Junwoo Yoon and Taegyu Jin and Seok-Hyung Bae},
  title     = {{SketchingRelatedWork}: Finding and Organizing Papers through Inking a Node-Link Diagram},
  booktitle = {Adjunct Proceedings of the 36th Annual ACM Symposium on User Interface Software and Technology},
  series    = {UIST '23 Adjunct},
  year      = {2023},
  pages     = {1--3},
  publisher = {ACM},
  doi       = {10.1145/3586182.3616685},
}

@article{endert2012semantic,
  author    = {Alex Endert and Patrick Fiaux and Chris North},
  title     = {Semantic Interaction for Sensemaking: Inferring Analytical Reasoning for Model Steering},
  journal   = {IEEE Transactions on Visualization and Computer Graphics},
  volume    = {18},
  number    = {12},
  year      = {2012},
  pages     = {2879--2888},
  doi       = {10.1109/TVCG.2012.260},
}

@inproceedings{masson2024directgpt,
  author    = {Damien Masson and Sylvain Malacria and G{\'e}ry Casiez and Daniel Vogel},
  title     = {{DirectGPT}: A Direct Manipulation Interface to Interact with Large Language Models},
  booktitle = {Proceedings of the 2024 CHI Conference on Human Factors in Computing Systems},
  series    = {CHI '24},
  year      = {2024},
  pages     = {1--16},
  publisher = {ACM},
  doi       = {10.1145/3613904.3642462},
}

@article{wang2025vizta,
  author    = {Liangwei Wang and Zhan Wang and Shishi Xiao and Le Liu and Fugee Tsung and Wei Zeng},
  title     = {{VizTA}: Enhancing Comprehension of Distributional Visualization with Visual--Lexical Fused Conversational Interface},
  journal   = {Computer Graphics Forum},
  volume    = {44},
  number    = {3},
  year      = {2025},
  pages     = {e70110},
  doi       = {10.1111/cgf.70110},
}

@inproceedings{amershi2019guidelines,
  author    = {Saleema Amershi and Dan Weld and Mihaela Vorvoreanu and Adam Fourney and Besmira Nushi and Penny Collisson and Jina Suh and Shamsi Iqbal and Paul N. Bennett and Kori Inkpen and Jaime Teevan and Ruth Kikin-Gil and Eric Horvitz},
  title     = {Guidelines for Human--AI Interaction},
  booktitle = {Proceedings of the 2019 CHI Conference on Human Factors in Computing Systems},
  series    = {CHI '19},
  year      = {2019},
  pages     = {1--13},
  publisher = {ACM},
  doi       = {10.1145/3290605.3300233},
}

@inproceedings{wu2022aichains,
  author    = {Tongshuang Wu and Michael Terry and Carrie J. Cai},
  title     = {{AI Chains}: Transparent and Controllable Human--AI Interaction by Chaining Large Language Model Prompts},
  booktitle = {Proceedings of the 2022 CHI Conference on Human Factors in Computing Systems},
  series    = {CHI '22},
  year      = {2022},
  pages     = {1--22},
  publisher = {ACM},
  doi       = {10.1145/3491102.3517582},
}

@inproceedings{shen2026iai,
  author    = {Leixian Shen and Yifang Wang and Huamin Qu and Xing Xie and Haotian Li},
  title     = {Interaction-Augmented Instruction: Modeling the Synergy of Prompts and Interactions in Human-GenAI Collaboration},
  booktitle = {Proceedings of the 2026 CHI Conference on Human Factors in Computing Systems},
  series    = {CHI '26},
  year      = {2026},
  publisher = {ACM},
  doi       = {10.1145/3772318.3790505},
}

@inproceedings{tang2026semanticprompting,
  author    = {Xuxin Tang and Ibrahim Tahmid and Eric Krokos and Kirsten Whitley and Xuan Wang and Chris North},
  title     = {Semantic Prompting: Agentic Incremental Narrative Refinement through Spatial Semantic Interaction},
  booktitle = {Proceedings of the 2026 International Conference on Advanced Visual Interfaces},
  series    = {AVI '26},
  year      = {2026},
  pages     = {1--9},
  publisher = {ACM},
  doi       = {10.1145/3811427.3811444},
}

@inproceedings{russell1993cost,
  author    = {Daniel M. Russell and Mark J. Stefik and Peter Pirolli and Stuart K. Card},
  title     = {The Cost Structure of Sensemaking},
  booktitle = {Proceedings of the INTERACT '93 and CHI '93 Conference on Human Factors in Computing Systems},
  series    = {INTERCHI '93},
  year      = {1993},
  pages     = {269--276},
  publisher = {ACM},
  doi       = {10.1145/169059.169209},
}

@inproceedings{pirolli2005sensemaking,
  author    = {Peter Pirolli and Stuart Card},
  title     = {The Sensemaking Process and Leverage Points for Analyst Technology as Identified Through Cognitive Task Analysis},
  booktitle = {Proceedings of the International Conference on Intelligence Analysis},
  volume    = {5},
  year      = {2005},
  pages     = {2--4},
}

@inproceedings{wright2006sandbox,
  author    = {William Wright and David Schroh and Pascale Proulx and Alex Skaburskis and Brian Cort},
  title     = {The Sandbox for Analysis: Concepts and Methods},
  booktitle = {Proceedings of the SIGCHI Conference on Human Factors in Computing Systems},
  series    = {CHI '06},
  year      = {2006},
  pages     = {801--810},
  publisher = {ACM},
  doi       = {10.1145/1124772.1124890},
}

@article{stasko2008jigsaw,
  author    = {John Stasko and Carsten G{\"o}rg and Zhicheng Liu},
  title     = {{Jigsaw}: Supporting Investigative Analysis through Interactive Visualization},
  journal   = {Information Visualization},
  volume    = {7},
  number    = {2},
  year      = {2008},
  pages     = {118--132},
  doi       = {10.1057/palgrave.ivs.9500180},
}

@inproceedings{bier2008entity,
  author    = {Eric A. Bier and Stuart K. Card and John W. Bodnar},
  title     = {Entity-Based Collaboration Tools for Intelligence Analysis},
  booktitle = {2008 IEEE Symposium on Visual Analytics Science and Technology (VAST)},
  year      = {2008},
  pages     = {99--106},
  publisher = {IEEE},
  doi       = {10.1109/VAST.2008.4677362},
}

@inproceedings{andrews2010spacetothink,
  author    = {Christopher Andrews and Alex Endert and Chris North},
  title     = {Space to Think: Large High-Resolution Displays for Sensemaking},
  booktitle = {Proceedings of the SIGCHI Conference on Human Factors in Computing Systems},
  series    = {CHI '10},
  year      = {2010},
  pages     = {55--64},
  publisher = {ACM},
  doi       = {10.1145/1753326.1753336},
}

@inproceedings{andrews2012analystworkspace,
  author    = {Christopher Andrews and Chris North},
  title     = {Analyst's Workspace: An Embodied Sensemaking Environment for Large, High-Resolution Displays},
  booktitle = {2012 IEEE Conference on Visual Analytics Science and Technology (VAST)},
  year      = {2012},
  pages     = {123--131},
  publisher = {IEEE},
  doi       = {10.1109/VAST.2012.6400559},
}

@inproceedings{jiang2023graphologue,
  author    = {Peiling Jiang and Jude Rayan and Steven P. Dow and Haijun Xia},
  title     = {{Graphologue}: Exploring Large Language Model Responses with Interactive Diagrams},
  booktitle = {Proceedings of the 36th Annual ACM Symposium on User Interface Software and Technology},
  series    = {UIST '23},
  year      = {2023},
  pages     = {1--20},
  publisher = {ACM},
  doi       = {10.1145/3586183.3606737},
}

@inproceedings{suh2023sensecape,
  author    = {Sangho Suh and Bryan Min and Srishti Palani and Haijun Xia},
  title     = {{Sensecape}: Enabling Multilevel Exploration and Sensemaking with Large Language Models},
  booktitle = {Proceedings of the 36th Annual ACM Symposium on User Interface Software and Technology},
  series    = {UIST '23},
  year      = {2023},
  pages     = {1--18},
  publisher = {ACM},
  doi       = {10.1145/3586183.3606756},
}

@inproceedings{kang2023synergi,
  author    = {Hyeonsu B. Kang and Sherry Tongshuang Wu and Joseph Chee Chang and Aniket Kittur},
  title     = {{Synergi}: A Mixed-Initiative System for Scholarly Synthesis and Sensemaking},
  booktitle = {Proceedings of the 36th Annual ACM Symposium on User Interface Software and Technology},
  series    = {UIST '23},
  year      = {2023},
  pages     = {1--19},
  publisher = {ACM},
  doi       = {10.1145/3586183.3606759},
}

@inproceedings{choe2026crosslit,
  author    = {Kiroong Choe and Eunhye Kim and Min-Hyeong Kim and Suyeon Hwang and Sangwon Park and Nam Wook Kim and Jinwook Seo},
  title     = {{CrossLit}: Connecting Visual and Textual Sensemaking for Literature Review},
  booktitle = {Proceedings of the 2026 CHI Conference on Human Factors in Computing Systems},
  series    = {CHI '26},
  year      = {2026},
  articleno = {407},
  numpages  = {30},
  publisher = {ACM},
  doi       = {10.1145/3772318.3791418},
}

@inproceedings{peng2025morae,
  author    = {Yi-Hao Peng and Dingzeyu Li and Jeffrey P. Bigham and Amy Pavel},
  title     = {{Morae}: Proactively Pausing UI Agents for User Choices},
  booktitle = {Proceedings of the 38th Annual ACM Symposium on User Interface Software and Technology},
  series    = {UIST '25},
  year      = {2025},
  pages     = {1--14},
  publisher = {ACM},
  doi       = {10.1145/3746059.3747797},
}

@inproceedings{huq2025cowpilot,
  author    = {Faria Huq and Zora Zhiruo Wang and Frank F. Xu and Tianyue Ou and Shuyan Zhou and Jeffrey P. Bigham and Graham Neubig},
  title     = {{CowPilot}: A Framework for Autonomous and Human--Agent Collaborative Web Navigation},
  booktitle = {Proceedings of the 2025 Conference of the Nations of the Americas Chapter of the Association for Computational Linguistics: Human Language Technologies (System Demonstrations)},
  year      = {2025},
  pages     = {163--172},
  publisher = {Association for Computational Linguistics},
  doi       = {10.18653/v1/2025.naacl-demo.17},
  url       = {https://aclanthology.org/2025.naacl-demo.17/},
}

@inproceedings{yang2025researstudio,
  author    = {Linyi Yang and Yixuan Weng},
  title     = {{ResearStudio}: A Human-intervenable Framework for Building Controllable Deep Research Agents},
  booktitle = {Proceedings of the 2025 Conference on Empirical Methods in Natural Language Processing: System Demonstrations},
  year      = {2025},
  pages     = {896--905},
  publisher = {Association for Computational Linguistics},
  doi       = {10.18653/v1/2025.emnlp-demos.69},
  url       = {https://aclanthology.org/2025.emnlp-demos.69/},
}

@inproceedings{cheng2026mapping,
  author    = {Ruijia Cheng and Jenny T. Liang and Eldon Schoop and Jeffrey Nichols},
  title     = {Mapping the Design Space of User Experience for Computer Use Agents},
  booktitle = {Proceedings of the 31st International Conference on Intelligent User Interfaces},
  series    = {IUI '26},
  year      = {2026},
  pages     = {646--662},
  publisher = {ACM},
  doi       = {10.1145/3742413.3789132},
}

@inproceedings{ye2025scholarmate,
  author    = {Runlong Ye and Patrick Lee and Matthew Varona and Oliver Huang and Carolina Nobre},
  title     = {{ScholarMate}: A Mixed-Initiative Tool for Qualitative Knowledge Work and Information Sensemaking},
  booktitle = {Adjunct Proceedings of the 4th Annual Symposium on Human-Computer Interaction for Work},
  series    = {CHIWORK '25 Adjunct},
  year      = {2025},
  pages     = {1--7},
  publisher = {ACM},
  doi       = {10.1145/3707640.3731913},
}

@inproceedings{ma2025gardenofpapers,
  author    = {Donghyeok Ma and Hanbee Jang and Joon Hyub Lee and Seok-Hyung Bae},
  title     = {Garden of Papers: Finding, Reading, and Organizing Research Papers in a Visual, Integrated, and Flexible Workspace},
  booktitle = {Proceedings of the 38th Annual ACM Symposium on User Interface Software and Technology},
  series    = {UIST '25},
  year      = {2025},
  pages     = {1--15},
  publisher = {ACM},
  doi       = {10.1145/3746059.3747637},
}

@inproceedings{vaithilingam2024dynavis,
  author    = {Priyan Vaithilingam and Elena L. Glassman and Jeevana Priya Inala and Chenglong Wang},
  title     = {{DynaVis}: Dynamically Synthesized UI Widgets for Visualization Editing},
  booktitle = {Proceedings of the 2024 CHI Conference on Human Factors in Computing Systems},
  year      = {2024},
  articleno = {985},
  numpages  = {17},
  doi       = {10.1145/3613904.3642639}
}

@misc{heptabase2026,
  author       = {{Heptabase}},
  title        = {Heptabase: An Intelligent, Visual Knowledge Base},
  year         = {2026},
  howpublished = {Product website},
  url          = {https://heptabase.com/},
  note         = {Accessed 2026-07-30}
}

@misc{miro2026canvascontent,
  author       = {{Miro}},
  title        = {Ways to Add Content to a Miro Board},
  year         = {2026},
  howpublished = {Miro Help Center},
  url          = {https://help.miro.com/hc/en-us/articles/360017731233-Ways-to-add-content},
  note         = {Updated 2026-02-06; accessed 2026-07-30}
}

@inproceedings{liu2026amiwrite,
  author    = {Ziyi Liu and Yuzhao Chen and Haoyu Ji and Runlin Duan and Zhengzhe Zhu and Xiyun Hu and Kylie A. Peppler and Karthik Ramani},
  title     = {{AmIWrite}: Exploring Scalable One-on-One Handwriting-Based Tutoring for Mathematical Problem-Solving with an {LLM}-Powered {AI} Tutor},
  booktitle = {Proceedings of the 2026 CHI Conference on Human Factors in Computing Systems},
  series    = {CHI '26},
  year      = {2026},
  articleno = {71},
  numpages  = {28},
  publisher = {ACM},
  doi       = {10.1145/3772318.3790935}
}

@inproceedings{yen2024memolet,
  author    = {Ryan Yen and Jian Zhao},
  title     = {{Memolet}: Reifying the Reuse of User-{AI} Conversational Memories},
  booktitle = {Proceedings of the 37th Annual ACM Symposium on User Interface Software and Technology},
  series    = {UIST '24},
  year      = {2024},
  pages     = {1--22},
  publisher = {Association for Computing Machinery},
  address   = {New York, NY, USA},
  doi       = {10.1145/3654777.3676388},
  url       = {https://doi.org/10.1145/3654777.3676388}
}

@misc{payandeh2026thinkink,
  author        = {Mohammad Hasan Payandeh and Daniel Vogel and Jian Zhao},
  title         = {{Thinkink}: {2D} Spatial Ink-native Interaction with {LLMs}},
  year          = {2026},
  eprint        = {2607.21468},
  archivePrefix = {arXiv},
  primaryClass  = {cs.HC},
  doi           = {10.48550/arXiv.2607.21468},
  url           = {https://arxiv.org/abs/2607.21468}
}

@misc{microsoft2026huabu,
  author       = {{Microsoft}},
  title        = {{Huabu}: Where You and Your Agents Think Together},
  year         = {n.d.},
  howpublished = {Project website},
  url          = {https://microsoft.github.io/Huabu/},
  note         = {Accessed 2026-09-10}
}

@misc{reactteam2026react,
  author       = {{React Team}},
  title        = {Quick Start -- {React}},
  year         = {n.d.},
  howpublished = {Online documentation},
  url          = {https://react.dev/learn},
  note         = {Accessed 2026-09-10}
}

@misc{xyflow2026reactflow,
  author       = {{xyflow}},
  title        = {Quick Start -- {React Flow}},
  year         = {n.d.},
  howpublished = {Online documentation},
  url          = {https://reactflow.dev/learn},
  note         = {Accessed 2026-09-10}
}

@misc{ruiz2026perfectfreehand,
  author       = {Steve Ruiz},
  title        = {{perfect-freehand}},
  year         = {n.d.},
  howpublished = {Software repository},
  url          = {https://github.com/steveruizok/perfect-freehand},
  note         = {Accessed 2026-09-10}
}

@misc{myscript2026iink,
  author       = {{MyScript}},
  title        = {{MyScript IInk Batch Mode}},
  year         = {n.d.},
  howpublished = {OpenAPI specification for the MyScript iink REST recognition service},
  url          = {https://cloud.myscript.com/api/v4.0/iink/batch/api-docs},
  note         = {Accessed 2026-09-10}
}

@misc{fastifyteam2026fastify,
  author       = {{The Fastify team}},
  title        = {Introduction -- {Fastify}},
  year         = {n.d.},
  howpublished = {Online documentation},
  url          = {https://fastify.dev/docs/latest/},
  note         = {Accessed 2026-09-10}
}

@misc{zedindustries2026acp,
  author       = {{Zed Industries, Inc. and contributors}},
  title        = {Introduction -- {Agent Client Protocol}},
  year         = {n.d.},
  howpublished = {Online protocol documentation},
  url          = {https://agentclientprotocol.com/get-started/introduction},
  note         = {Accessed 2026-09-10}
}

@misc{modelcontextprotocol2026mcp,
  author       = {{Model Context Protocol}},
  title        = {What is the {Model Context Protocol (MCP)}?},
  year         = {n.d.},
  howpublished = {Online protocol documentation},
  url          = {https://modelcontextprotocol.io/docs/getting-started/intro},
  note         = {Accessed 2026-09-10}
}

@misc{openai2026gpt56sol,
  author       = {{OpenAI}},
  title        = {{GPT-5.6 Sol}},
  year         = {n.d.},
  howpublished = {OpenAI API model documentation},
  url          = {https://developers.openai.com/api/docs/models/gpt-5.6-sol},
  note         = {Accessed 2026-09-10}
}

\appendix

\end{document}